\documentclass{article}
 
\usepackage[a4paper, total={6in, 9in}]{geometry}
\usepackage{graphicx}
\usepackage{amssymb}
\usepackage{mathtools}
\usepackage{amsmath}
\usepackage{appendix}
\usepackage{comment}
\usepackage{amssymb}
\usepackage{physics}
\usepackage{color}
\usepackage{cleveref}
\usepackage{cases}
\usepackage{empheq}
\usepackage{multirow}
\usepackage{arydshln}
\usepackage{booktabs}
\usepackage{authblk}
\usepackage{cite}

\newcommand{\inert}{\text{in}}
\newcommand{\im}{\text{im}}
\newcommand{\corr}{\text{corr}}
\newcommand{\comp}{\text{comp}}
\newcommand{\F}{\mathcal{F}}

\DeclareMathOperator{\sgn}{sgn}
\DeclareMathOperator{\sinc}{sinc}
\DeclareMathOperator{\sinch}{sinch}
\newcommand{\intoinf}{\int_0^{\infty}}
\newcommand{\CM}{\text{CM}}
\newcommand{\DR}{\text{DR}}
\newcommand{\intinf}{\int_{-\infty}^{\infty}}
\newcommand{\odd}{\text{odd}}

\newcommand{\PA}{\text{PA}}
\newcommand{\lin}{\text{lin}}

\DeclareMathOperator{\arccosh}{arccosh}

\numberwithin{equation}{section}

\title{Connecting the circular and drifted Rindler Unruh effects}

\author{Leo J. A. Parry\thanks{leo.parry@nottingham.ac.uk}\ }
\author{Jorma Louko\thanks{jorma.louko@nottingham.ac.uk}}

\affil{School of Mathematical Sciences,
University of Nottingham,\\
Nottingham NG7 2RD, UK}

\date{September 2024. Revised November 2024.\\[2ex]
{\small Published in Phys.\ Rev.\ D \textbf{111}, 025012 (2025).}}

\begin{document}

\maketitle

\begin{abstract}

In Minkowski spacetime quantum field theory, each stationary motion is associated with an effective, energy-dependent notion of temperature, which generalises the familiar Unruh temperature of uniform linear acceleration. 
Motivated by current experimental interest in circular motion, 
we analyse the effective temperature for drifted Rindler motion, generated by a boost and a spacelike translation (drift), and the way in which drifted Rindler motion can be smoothly 
(and in fact real analytically)
deformed to circular motion through a third type of motion known as parator. 
For an Unruh-DeWitt detector coupled linearly to a massless scalar field in $2+1$ and $3+1$ spacetime dimensions, we establish analytic results in the limits of large gap, small gap and large drift speed. For fixed proper acceleration, the drifted Rindler temperature remains bounded in the large gap limit, in contrast to the circular motion temperature, which can be arbitrarily large in this limit. Finally, in 
$2+1$ 
dimensions, we trace the vanishing of the circular motion temperature in the small gap limit to the weak decay of the Wightman function, and we show that, among all types of stationary motion in all dimensions, this phenomenon is unique to $2+1$ dimensions and therein to circular and parator motion.
\end{abstract}

\section{Introduction}

The Unruh effect \cite{Fulling1973,Davies1975,Unruh1976,Fulling:2014} 
predicts that a uniformly linearly accelerated observer with proper acceleration $a$ reacts to the 
Minkowski vacuum of a relativistic quantum field as if the vacuum 
were a thermal state with the Unruh temperature
\begin{equation} \label{Unruh temp}
    T_U = \frac{\hbar a}{2\pi c k_B}.
\end{equation}
This effect is a consequence of the observer-dependence of the notion of a ``particle'' in relativistic quantum field theory, in flat and curved spacetimes \cite{birrell,Fullingbook}. 
Related predictions include the Hawking effect \cite{Hawking:1975vcx} and the cosmological particle creation \cite{Parker:1969au} from which the present-day structure of the Universe may originate~\cite{Mukhanov:2007zz}. 

In the standard setting of a relativistic quantum field, the Unruh effect is very small. The acceleration necessary to register a temperature of $1$K is approximately $2.4 \times 10^{20}$m/s$^{2}$, and the effect has not been experimentally verified. 
The prospects to observe the effect are however better in analogue spacetime
systems \cite{Unruh1981,Liberati,HeliumUniverse} where condensed matter excitations simulate a relativistic quantum field, 
but with the speed of light replaced by the speed of sound thereby raising the Unruh temperature \eqref{Unruh temp} by several orders of magnitude. In this setting, the classical mode conversion underlying the Hawking and Unruh effects has been observed \cite{Weinfurtner:2010nu,Leonhardt:2017lwm}. 
The experimental prospects in a tabletop system are further enhanced by considering circular rather than linear acceleration, where the accelerating trajectory can be kept in a finite spatial volume for an arbitrarily long time \cite{Retzker:2007vql,BEC,Lochan:2019osm,Bunney:2023ude}, and where the lack of a condensed matter relativistic time dilation can be accounted for at the data analysis stage \cite{Retzker:2007vql,BEC,Bierman2020}. 
An earlier context where the circular motion Unruh effect has been considered is the depolarisation of electron beams in accelerator storage rings 
\cite{BellLeinaas,Bell:1986ir,Costa:1994yx,Guimaraes:1998jf,Leinaas:1998tu,Unruh:1998gq}. 
Other work on the theory of the circular motion Unruh effect includes 
\cite{Denardo:1978dj,Letaw:1979wy,LetawStationary,Takagi:1986kn,Davies:1996ks,Korsbakken:2004bv,Chowdhury:2019set,Good:2020hav,Unruh:2022gso,Bunney:2023vyj,Bunney:2023ipk,Bunney:2024qrh}.

A subtlety with circular acceleration is that the linear acceleration Unruh temperature formula \eqref{Unruh temp} is no longer exact, 
and the circular acceleration effect in fact cannot be described in terms of a density matrix with a single `temperature' parameter \cite{Denardo:1978dj,Letaw:1979wy,Davies:1996ks}. This is 
in contrast to the linear acceleration effect, 
which has a well-known description in terms of a Bogoliubov transformation between the Minkowski and Rindler vacua~\cite{Unruh1976}. Nevertheless, an effective temperature for circular acceleration can be introduced in terms of the excitations and de-excitations of a local quantum system following the accelerated worldline, and although this effective temperature depends on the internal energy spacing of the system, it is in broad agreement with the Unruh temperature formula \eqref{Unruh temp} over most of the parameter space \cite{Unruh:1998gq,Good:2020hav,Bierman2020}. The effective temperature therefore provides a useful quantifier of the acceleration effects. 

That being said, a puzzle with the circular acceleration effective temperature is that it is much smaller than the linear acceleration Unruh temperature \eqref{Unruh temp} when the effective spacetime dimension is $2+1$ (as it is in the Bose-Einstein condensate and superfluid helium analogue spacetime proposals in \cite{BEC,Bunney:2023ude}) and the internal energy spacing of the accelerating system is small~\cite{Bierman2020}. The purpose of this paper is to investigate the deviation from the linear acceleration effective temperature for small internal energy spacings from the broader perspective of observers in arbitrary types of stationary motion \cite{Good:2020hav,LetawStationary,Takagi:1986kn,Korsbakken:2004bv}. In particular, as the circular acceleration Killing vector is a linear combination of a spatial rotation and a time translation, circular acceleration can be viewed as dual to the drifted Rindler motion~\cite{LetawStationary,Takagi:1986kn}, whose Killing vector is a linear combination of a boost and a spatial translation. We show how these two types of motions can be smoothly deformed to each other through a type of motion known as parator motion, which is generated by a null rotation and a timelike translation \cite{LetawStationary,Good:2020hav}, 
and we observe that the deformation can be viewed as the unique (real) analytic continuation in the parameters of the motion. 
We further show how the effective temperature undergoes qualitative changes in this deformation, particularly in the regime where the detector's internal energy spacing is small. We trace the smallness of the circular motion effective temperature for small internal energy spacings to the weak decay of the Wightman function along the detector's trajectory, and we show that, among all types of stationary motion in all dimensions, this phenomenon is unique to $2+1$ dimensions and therein to circular and parator motion.

We work in a technical setting where the quantum field is a real massless scalar field in Minkowski spacetime of dimension $2+1$ or higher, prepared in its Minkowski vacuum state. We probe the field with an Unruh-DeWitt (UDW) detector, a pointlike two-level system coupled linearly to the scalar field \cite{Unruh1976,DeWitt1979}, and we treat the coupled system to leading order in perturbation theory. This model captures the essentials of the interaction between atomic orbitals and the electromagnetic 
field~\cite{Martin-Martinez:2012ysv,Alhambra:2013uja}. 
As the detector's worldline is by assumption stationary, and the Minkowski vacuum is Poincar\'e invariant, the coupled system is invariant under time translations along the detector's worldline, and we can consider the detector's transition probability per unit time \cite{Lin:2006jw,Benatti:2004ee,DeBievre:2006pys,waitingforunruh}. We can then characterise the detector's response by an effective temperature, defined by fitting the ratio of the excitation and de-excitation rates to the exponential formula that this ratio obeys in a genuine thermal ensemble \cite{BellLeinaas,Takagi:1986kn,Unruh:1998gq}. 
The resulting temperature, which we call the detailed balance temperature, depends on the detector's energy gap, and the puzzle for circular acceleration in $2+1$ dimensions is that the detailed balance temperature goes to zero linearly as the gap goes to zero \cite{Bierman2020}. It is in terms of the detailed balance temperature that we shall analyse the status of $(2+1)$-dimensional circular motion among stationary motions in all dimensions, $2+1$ and higher. 

The key mathematical observation in this technical setting is that the detector's response function is the Fourier transform of the pullback of the field's Wightman distribution to the detector's worldline, and one therefore expects that the small gap behaviour of the response is determined by the large time decay of the Wightman distribution. The decay depends both on the spacetime dimension and the detector's trajectory. We show that in almost all cases the decay is indeed so strong that the detailed balance temperature remains finite in the small gap limit. The only exceptions occur in $2+1$ dimensions, and therein only for two types of accelerated motion. For circular motion, the decay of the Wightman function is proportional to the inverse of the proper time, and the detailed balance temperature falls off linearly in the gap when the gap is small. For parator motion, the decay of the Wightman function is proportional to the inverse square of the proper time, and the detailed balance temperature falls off as the inverse of the logarithm of the gap when the gap is small. 

As an intermediate step in the analysis, we express the stationary response function in an arbitrary dimension as a formula in which the contribution from the distributional part of the Wightman function has been recast as a polynomial in the gap, while the remaining contribution is the Fourier transform of a smooth function. The stationarity of the detector's trajectory allows the split to be performed via a simple Laurent expansion around the distributional singularities in any dimension. 
For motion that is not necessarily stationary, corresponding expressions for the instantaneous transition rate in dimensions up to $5+1$ have been given in~\cite{Loukoregspatialprofile,Hodgkinson_2012}. 

For the connection between circular motion and drifted Rindler motion, the key mathematical observation is that the deformation of these two types of motion to each other through parator motion is entirely 
smooth, and in fact real analytic in the parameters of the motion. 
This relationship has been described previously in terms of limits in $3+1$ dimensions in~\cite{Takagi:1986kn,Good:2020hav}, however we describe this deformation in terms of the underlying two-parameter family of Killing vectors, with the parator Killing vectors as a one-parameter subfamily separating circular motion from drifted Rindler motion. In particular, in any dimension, $2+1$ and higher, parator motion can be understood as the ultrarelativistic limit of circular motion as described in the Lorentz frame adapted to the circular motion \cite{Unruh:1998gq,Bierman2020}, 
and the same holds for the high drift speed limit of drifted Rindler motion.

For drifted Rindler motion, we also show that in both $2+1$ and $3+1$ dimensions the drift speed has a modest heating effect relative to Rindler motion in the large gap regime, and we expect the same to hold in all dimensions. In the small gap regime, the drift speed has a modest cooling effect in $3+1$ dimensions, but a stronger cooling effect in $2+1$ dimensions, where the detailed balance temperature approaches zero as the drift speed approaches unity. 

This paper is structured as follows. In Section \ref{DR CM Geometry}, we review stationary worldlines in Minkowski spacetime and present the smooth deformation of circular motion to drifted Rindler motion through parator motion. 
In Section \ref{UDW model}, we introduce the UDW detector model and its response in stationary motion, expressing the response as an integral formula without distributional singularities. 
Section \ref{Temp asymptotics} gives a detailed analysis of drifted Rindler motion in $2+1$ and $3+1$ dimensions in several asymptotic regimes, with comparison to the corresponding results in circular motion~\cite{Bierman2020}, and Section \ref{Exceptionality section} establishes the small gap exceptionality that occurs in circular motion and parator motion in $2+1$ dimensions. 
Section \ref{sec:conclusions} presents a summary and concluding remarks. 
Technical results are deferred to five appendices. 

We use units in which $\hbar=k_B = c =1$. 
We work in $d$-dimensional Minkowski spacetime with $d\ge3$, 
with standard Minkowski coordinates $(t, \ x^1, \ x^2, \ \dots, \ x^{d-1} ) = (t, \mathbf{x})$, in which the Minkowski metric $\eta$ reads 
$ds^2 = - dt^2 + {(dx^1)}^2 + \cdots + {(dx^{d-1})}^2$. 
Spacetime points are denoted by sans serif letters. 
In asymptotic formulae, 
$f(x)=O(x)$ denotes that $f(x)/x$ is bounded in the limit of interest, 
and $f(x) = o(x)$ denotes that $f(x)/x$ tends to zero in the limit of interest. The Heaviside theta function $\Theta(x)$ is defined as
\begin{equation}
    \Theta(x) = 
    \begin{cases}
        1 & \text{ for } x\geq0 \\
        0 & \text{ for } x<0,
    \end{cases}
\end{equation}
and the signum function $\sgn(x)$ is defined as
\begin{align}
    \sgn(x) = 
    \begin{cases}
        1 & \text{ for } x>0 \\
        -1 & \text{ for } x<0 \\
        0 & \text{ for } x=0.
    \end{cases}
\end{align}

\section{Drifted Rindler and Circular Worldlines} \label{DR CM Geometry}

We begin this section by reviewing the description of an arbitrary timelike worldline $\mathsf{x}^\mu(\tau)$ parametrised by proper time $\tau$ in four-dimensional Minkowski spacetime in the context of the tetrad formalism \cite{LetawStationary}. The tetrad formalism replaces the usual coordinate basis $\{\partial_\mu\}$ of the tangent bundle with a more general local basis of each open set of an open cover of the spacetime manifold. Such a local basis is a set of four linearly-independent vector fields $\{V_a\}$ that has the following expansion with respect to the coordinate basis $\{\partial_\mu\}$
\begin{equation}
    V_a = V_a^\mu \partial_\mu.
\end{equation}
We further impose the local orthonormality condition on this set
\begin{equation} 
    V_{a \mu} V^\mu_b = \eta_{ab},
\end{equation}
which is satisfied in each relevant open set. This basis is referred to as an orthonormal tetrad. For more details on the tetrad approach to general relativity, see for example \cite{Wald:1984rg}.

In order to describe a timelike worldline $\mathsf{x}^\mu(\tau)$ parameterised by proper time $\tau$, we construct an orthonormal tetrad $\{V_a^\mu(\tau)\}$ that is a basis for the tangent space at each $\tau$ along the worldline. The first element of the tetrad $V_0^\mu(\tau)$ is the four-velocity $\dot{\mathsf{x}}^\mu (\tau)$ of the worldline and the remaining elements are found by applying the Gram-Schmidt process to the set $\{\dot{\mathsf{x}}^\mu,\ddot{\mathsf{x}}^\mu,\dddot{\mathsf{x}}^\mu,\ddddot{\mathsf{x}}^\mu\}$, consisting of the proper time derivatives of $\mathsf{x}^\mu$ up to fourth order. The derivatives of $V_a^\mu (\tau)$ with respect to proper time can be expressed as a linear combination of the elements of the tetrad, as
\begin{equation} \label{Frenet Serret}
    \dot{V_a^\mu}(\tau) = K_a^{ \ b}(\tau) V_b^\mu(\tau),
\end{equation}
where $K_{ab}(\tau)$ is an antisymmetric matrix given by
\begin{equation} \label{kab matrix}
K_{ab} =
    \begin{pmatrix}
        0 & a(\tau) & 0 & 0 \\
        -a(\tau) & 0 & b(\tau)  & 0 \\
        0 & -b(\tau) & 0 & \nu(\tau) \\
        0 & 0 & -\nu(\tau) & 0
    \end{pmatrix}.
\end{equation}
Hence, we can describe timelike worldlines as the solutions to the generalised Frenet-Serret equations, \eqref{Frenet Serret} and \eqref{kab matrix}, in terms of three curvature invariants; curvature $a(\tau)$, torsion $b(\tau)$ and hypertorsion $\nu(\tau)$. In this setting, the curvature is the proper acceleration of the worldline while the torsion and hypertorsion are the components of the proper angular velocity of the spatial frame $\{V^\mu_1, V^\mu_2, V^\mu_3\}$ in the planes spanned by $\{V_1^\mu, V_2^\mu\}$ and $\{V_2^\mu, V_3^\mu\}$, respectively. 

This construction is easily generalised to $d$-dimensional Minkowski spacetime for $d\ge2$, 
by applying the Gram-Schmidt orthogonalisation process to $d$th order. The resulting $d$-dimensional local orthonormal basis is referred to as a vielbein and it satisfies the generalised Frenet-Serret equations \eqref{Frenet Serret} where $K_{ab}$ is now a $d\times d$ antisymmetric matrix given in terms of $d-1$ curvature invariants. See \cite{bunney2023stationary} for more details. 
In particular, for $d=3$, there is curvature $a(\tau)$ and torsion $b(\tau)$ but no hypertorsion $\nu(\tau)$, 
and $K_{ab}$ is a $3\times 3$ matrix obtained by deleting the last row and the last column in~\eqref{kab matrix}. 

A worldline that is a solution to \eqref{Frenet Serret} with constant curvature invariants is called stationary. Equivalently, a stationary worldline can be defined as an orbit of a Killing vector field that is timelike and future-pointing in a neighbourhood of the worldline. Stationary worldlines have the property that the geodesic distance between any two points depends only on the difference in proper time between them. In four-dimensional Minkowski spacetime, stationary worldlines can be categorised into six basic families depending on the relative magnitudes of the curvature invariants \cite{LetawStationary,Good:2020hav,fewster2023quantum}. 
In the rest of this section, we describe in detail three of these families: 
circular motion, drifted Rindler motion, and the interpolating case of parator motion. 
We write the formulae in four spacetime dimensions, but as all three families have vanishing hypertorsion, the description also applies in three spacetime dimensions by dropping the last spatial coordinate in \eqref{CM worldlime}, \eqref{DR worldline} and~\eqref{PA worldline}. 

Firstly, circular motion can be parametrised by the radius $R$ and the orbital speed $v$ where $R>0$ and $0<v<1$. In an adapted Lorentz frame, the worldline can be written as
\begin{equation} \label{CM worldlime}
    \mathsf{x}_\CM (\tau) = \Big( \gamma \tau, \ R \cos\big(\tfrac{\gamma v}{R}\tau \big) , \ R \sin\big(\tfrac{\gamma v}{R}\tau \big) , \ 0 \Big),
\end{equation}
where $\tau$ is the proper time and $\gamma = (1-v^2)^{-1/2}$. Note that since $v$ is the speed of the worldline, $\gamma$ is the usual Lorentz factor and it is constant in $\tau$.  Circular motion has zero hypertorsion $\nu$, but it has non-zero proper acceleration $a$ and non-zero torsion $b$, such that $|b|>|a|$. The proper acceleration and torsion are given in terms of $R$ and $v$ as
\begin{equation} \label{CM acc tors}
    a = \frac{\gamma^2 v^2}{R} \ , \ \ \ \ \ \ \ \ \ \ b = \frac{\gamma^2v}{R},
\end{equation}
which implies that $v=\frac{a}{b}$. The corresponding Killing vector is 
\begin{equation} \label{CM KV}
    \xi_{\CM} = \gamma \partial_t + \frac{\gamma v}{R}\big(  x\partial_y - y\partial_x  \big),
\end{equation}
where we have chosen the normalisation such that the flow parameter is the proper time. 

Secondly, drifted Rindler motion is the combination of linear acceleration and a spacelike translation at constant speed in a transverse direction. We refer to this spacelike translation as a drift. In an adapted Lorentz frame, the drifted Rindler worldline can be written as
\begin{equation} \label{DR worldline}
    \mathsf{x}_\DR (\tau) = \Big( R \sinh\big(\tfrac{\gamma }{R}\tau \big), \ R \cosh\big(\tfrac{\gamma}{R}\tau \big) , \ \gamma v \tau , \ 0 \Big),
\end{equation}
where $\tau$ is the proper time, $R>0$ is the distance of closest approach to the origin in the adapted frame, $v$ is the drift speed satisfying $0<v<1$ and $\gamma = (1-v^2)^{-1/2}$. Note that the speed of the worldline in this frame equals 
$\sqrt{v^2 + (1-v^2)\tanh^2(\gamma \tau /R)}$, which depends on $\tau$ and equals the drift speed only at the moment of closest approach to the origin, $\tau=0$. Hence, $\gamma$ is the Lorentz factor only at $\tau=0$.

Like circular motion, drifted Rindler motion also has zero hypertorsion $\nu$, non-zero proper acceleration $a$ and non-zero torsion $b$, but now such that $|a|>|b|$. $a$ and $b$ are now given in terms of $R$ and $v$ as
\begin{equation} \label{DR acceleration torsion}
    a = \frac{\gamma^2 }{R} \ , \ \ \ \ \ \ \ \ \ \ b = \frac{\gamma^2v}{R},
\end{equation}
which implies that $v=\frac{b}{a}$. The corresponding Killing vector is
\begin{equation} \label{DR KV}
     \xi_\DR = \frac{\gamma}{R} \big( x\partial_t + t\partial_x \big) + \gamma v \partial_y,
\end{equation}
where we have 
again chosen the normalisation such that the flow parameter is the proper time. 

It was shown in \cite{Good:2020hav} that in the $v\rightarrow 1$ limit, both the circular and the drifted Rindler worldlines reduce to a third type of stationary worldline, referred to variously as parator motion, cusped motion or semi-cubical parabolic motion. In this paper, we refer to this motion as parator motion. It has zero hypertorsion $\nu$, but non-zero 
proper acceleration $a$ and non-zero torsion $b$ with equal magnitudes, $|a|=|b|$. 
In an adapted Lorentz frame, the parator worldline is
\begin{equation} \label{PA worldline}
   \mathsf{x}_\PA(\tau) =  \Big( \tau + \frac{1}{6}a^2\tau^3, \ \frac{1}{2}a\tau^2 , \ \frac{1}{6}a^2\tau^3 , \ 0 \Big),
\end{equation}
where $\tau$ is the proper time and $a>0$. 
The corresponding Killing vector is
\begin{equation} \label{PA KV}
    \xi_\PA = \partial_t + a\big(x\partial_t + t\partial_x - y\partial_x + x\partial_y \big), 
\end{equation}
which is a combination of a timelike translation and a null rotation with relative weights determined by the proper acceleration $a$.

We wish to observe here that parator motion is not just a limiting case of circular motion and drifted Rindler motion: the parator one-parameter family smoothly connects the circular motion two-parameter family to the drifted Rindler two-parameter family as a one-dimensional surface in a two-dimensional parameter space that contains all three types of motion. This is conveniently seen in terms of the Killing vectors \eqref{CM KV}, \eqref{DR KV} and \eqref{PA KV} as follows.

Consider the circular motion Killing vector $\xi_\CM$  \eqref{CM worldlime}. Under the coordinate transformation
\begin{subequations}\label{boost transformation}
\begin{align} 
    t' &= \gamma (t-vy), \label{trans1} \\
    x' &= x-R,  \\
    y' &= \gamma (y-vt), \label{trans3}
\end{align}
\end{subequations}
which is a boost in $(t,y)$ and a translation in $x$, $\xi_\CM$ becomes
\begin{equation} \label{CM KV boosted}
\xi_\CM = \partial_{t'} - a(x'\partial_{t'} + t' \partial_{x'} ) -\frac{a}{v}(y'\partial_{x'} - x'\partial_{y'}), 
\end{equation}
where we have used \eqref{CM acc tors} to adopt $a$ and $v$ as the two independent parameters. 
Letting $(x',y') \rightarrow (-x',-y')$, which is a rotation by $\pi$ in $(x',y')$, 
and writing $a/v = b$, where $b$ is the circular motion torsion, 
$\xi_\CM$ becomes
\begin{equation} \label{CM KV boosted rotated}
\xi_\CM = \partial_{t'} + a(x'\partial_{t'} + t' \partial_{x'} ) -b(y'\partial_{x'} - x'\partial_{y'}), 
\end{equation}
where $0<a<b$ by construction. 

Consider then the drifted Rindler Killing vector \eqref{DR KV}. Applying the coordinate transformation \eqref{boost transformation}, with the relevant symbols now defined in \eqref{DR acceleration torsion}, 
brings $\xi_\DR$ to the form  
\begin{equation} \label{DR ultrarel KV}
  \xi_\DR = \partial_{t'} + a(x'\partial_{t'} + t' \partial_{x'} ) +av(y'\partial_{x'} - x'\partial_{y'}).
\end{equation}
Letting $y'\rightarrow -y'$, which is a reflection in $y'$, 
and writing $av = b$, where $b$ is the drifted Rindler torsion, 
$\xi_\DR$ becomes
\begin{equation} \label{DR ultrarel KV 2}
  \xi_\DR = \partial_{t'} + a(x'\partial_{t'} + t' \partial_{x'} ) -b (y'\partial_{x'} - x'\partial_{y'}), 
\end{equation}
where now $0<b<a$ by construction. 

Comparing \eqref{PA KV}, \eqref{CM KV boosted rotated}
and \eqref{DR ultrarel KV 2}, it is plain that the circular motion Killing vectors, the drifted Rindler Killing vectors and the parator Killing vectors form a smooth two-parameter family, given by \eqref{DR ultrarel KV 2} with $a>0$ and $b>0$, such that circular motion occurs for $a<b$, drifted Rindler motion occurs for $b<a$, and the two are joined by the parator one-parameter subfamily in which $a=b$. In particular, parator motion can be obtained, through the boosts described above, both as the ultrarelativistic $v\to1$ limit of circular motion in \eqref{CM worldlime} and as the corresponding $v\to1$ limit of drifted Rindler motion in \eqref{DR worldline} \cite{Good:2020hav}. 

We observe further that the Killing vector family \eqref{DR ultrarel KV 2} is not just smooth in $a$ and $b$ but (real) analytic in each. The same holds for the trajectory $\mathsf{x}(\tau)$ when written out in the Lorentz frame of \eqref{DR ultrarel KV 2} \cite{Good:2020hav}. 
The connection of circular motion and drifted Rindler motion can therefore be viewed as a (real) analytic continuation in $a$ and~$b$, 
and, by uniqueness of analytic continuation, it is the unique real analytic continuation. 
This observation extends to the spacetime interval $\Delta \mathsf{x}^2 (s) := \big( \mathsf{x}(s)-\mathsf{x}(0) \big)^2$ on the trajectory: adapting the formulas of \cite{Good:2020hav} to our notation, we have  
\begin{align}
- \frac{\Delta \mathsf{x}^2 (s)}{s^2}
= 
\begin{cases}
{\displaystyle 1 + a^2 \frac{\sinc^2 \bigl(\tfrac12 \sqrt{b^2-a^2} \, s \bigr) - 1}{a^2 - b^2}}
& \text{for}\ a<b \,,
\\[1ex]
{\displaystyle 1 + \tfrac{1}{12}a^2 s^2}
& \text{for}\ a=b \,,
\\[1ex]
{\displaystyle 1 + a^2 \frac{\sinch^2 \bigl(\tfrac12 \sqrt{a^2-b^2} \, s \bigr) - 1}{a^2 - b^2}}
& \text{for}\ a>b \,,
\end{cases}
\end{align}
where 
\begin{subequations}
\begin{align}
\sinc(z) &= \begin{cases}
{\displaystyle \frac{\sin z}{z}} & \text{for}\ z\ne0 \,,
\\
1 & \text{for}\ z=0 \,,
\end{cases}
\\
\sinch(z) &= \begin{cases}
{\displaystyle \frac{\sinh z}{z}} & \text{for}\ z\ne0 \,,
\\
1 & \text{for}\ z=0 \,,
\end{cases}
\end{align}
\end{subequations}
from which the analyticity in $a$ and $b$ is clear. 

Whether this sense of analytic continuation in $a$ and $b$ extends to the detector response function \eqref{stationary RF} that we study in the rest of this paper would require a more detailed analysis of the integral in~\eqref{stationary RF}, especially across the parator subfamily $a=b$, where the integrand's falloff properties undergo a qualitative change. 
We shall not address this question in this paper.

\section{Field-detector model} \label{UDW model}
In this section, we first review the linear interaction of a UDW detector with a real massless scalar field in Minkowski spacetime of dimension $d\ge3$. We then focus on stationary trajectories and present the detector's response as an integral formula without distributional singularities for any $d\ge3$, 
deferring details to Appendix \ref{RF derivation} for $d\ge5$. 
These integral formulae provide the starting point of the asymptotic analyses in the later sections.

\subsection{Unruh-DeWitt detector}
We begin by describing the UDW detector model \cite{Unruh1976,DeWitt1979}. 
We take the detector to be a two-level quantum system described by a Hamiltonian $H_D$ with two orthonormal energy eigenstates $\ket{0}_D$ and $\ket{1}_D$ with energy eigenvalues $0$ and $E$, respectively. The energy difference between the two levels is referred to as the energy gap. If $E>0$, then $\ket{0}_D$ is the ground state and $\ket{1}_D$ is the excited state. If $E<0$, then the roles are reversed.

 We use this system to probe a real massless scalar field $\phi(\mathsf{x})$ prepared initially in the state $\ket{\Psi}$. We assume that the Wightman function $\mathcal{G}(\mathsf{x},\mathsf{x}') = \expval{\phi(\mathsf{x})\phi(\mathsf{x}')}{\Psi}$ is a distribution of Hadamard type in the coincidence limit $\mathsf{x}\rightarrow \mathsf{x}'$ \cite{Decanini:2005eg, Kay:1988mu}.
 
 We couple the detector linearly to the field via the monopole moment operator $\mu(\tau)$ and allow the detector to move along a given worldline $\mathsf{x}(\tau)$ parametrised by proper time $\tau$. We can therefore write the interaction Hamiltonian as
\begin{equation}
    H_{\text{int}} = \lambda \chi(\tau) \mu(\tau)\phi\big(\mathsf{x}(\tau)\big),
\end{equation}
where $\lambda$ is a coupling constant and $\phi\big(\mathsf{x}(\tau)\big)$ is the value of the field pulled back to the worldline of the detector. The switching function $\chi(\tau) \in C^\infty_0(\mathbb{R})$ specifies how the interaction is turned on and off. This is not the most general field-detector coupling since it is smeared only in time, but not in space.  It is for this reason that this particular theory describes the interaction between the field and a \emph{pointlike} detector. For more general treatments using spatially smeared detectors, see e.g. \cite{Martin-Martinez:2020pss, Martin-Martinez:2020lul}.

Before the interaction is switched on, if the detector is in the state $\ket{0}_D$ and the field is in the state $\ket{\Psi}$, then after the interaction has taken place, there is a finite probability of finding the detector in the state $\ket{1}_D$. To first order in perturbation theory, this probability is proportional to the response function $\F_\chi (E)$ with a proportionality factor that depends only on the coupling constant and the internal structure of the detector. The response function $\F_\chi (E)$ is given by
\begin{equation} \label{Resp Funct Switch}
    \mathcal{F}_\chi(E) = \int_{-\infty}^{\infty} d\tau' \int_{-\infty}^{\infty} d\tau'' e^{-iE(\tau'-\tau'')} \chi(\tau')\chi(\tau'') \mathcal{W}(\tau',\tau''),
\end{equation}
where $\mathcal{W}(\tau,\tau') := \mathcal{G}\big( \mathsf{x}(\tau), \mathsf{x}(\tau') \big)$ is the Wightman function in the state $\ket{\Psi}$ pulled back to the worldline of the detector. 

For a real massless scalar field in the Minkowski vacuum $\ket{0}$ in $d$-dimensional Minkowski spacetime with $d\ge3$, 
the Wightman function $\mathcal{W}(\tau,\tau')$ is usually represented with an $i\epsilon$-regulator in the following way
\begin{equation} \label{Wightman d dim}
 \mathcal{W}(\tau,\tau') = \frac{\Gamma(d/2-1)}{4\pi^{d/2}\big[ (\mathbf{x}-\mathbf{x}')^2 - (t-t'-i\epsilon)^2 \big]^{(d-2)/2}},
\end{equation}
where $\mathbf{x} = \mathbf{x}(\tau)$, $\mathbf{x}' = \mathbf{x}(\tau')$, $t = t(\tau)$ and $t'=t(\tau')$, and where the distributional limit $\epsilon\rightarrow 0^+$ is understood. For odd $d$, the branch of the fractional power for timelike separations is specified by analytic continuation from spacelike separations, with the outcome that the denominator has the phases $i^{d-2}$ or $(-i)^{d-2}$ when $t-t'>0$ or $t-t'<0$, respectively.

\subsection{Stationary response function} \label{Section stationary RF}

We now specialise to a detector in stationary motion and assume that the field has been prepared in the Minkowski vacuum. In this case, the stationarity of the detector's worldline implies that the Wightman function depends only on the difference in proper time between any two points i.e. $\mathcal{W}(\tau,\tau') = \mathcal{W}(\tau-\tau',0)$. In the limit of long interaction duration, while keeping the coupling constant $\lambda$ so small that first-order perturbation theory is still valid, the detector's transition probability per unit time is then proportional to the stationary response function, given by \cite{Unruh1976,DeWitt1979,birrell,waitingforunruh}
\begin{equation} \label{stationary RF}
    \mathcal{F}(E)  =  \intinf ds \, e^{-iEs} \, \mathcal{W}(s),
\end{equation}
where $\mathcal{W}(s) := \mathcal{W}(s,0)$ is given by
\begin{equation} \label{wightman d dimensions stationary}
    \mathcal{W}(s) = \frac{\Gamma(d/2-1)}{4\pi^{d/2}\big[ \Delta \mathsf{x}^2(s-i\epsilon) \big]^{(d-2)/2}},
\end{equation}
where $\Delta \mathsf{x}^2 (s) := \big( \mathsf{x}(s)-\mathsf{x}(0) \big)^2$, which we refer to as the spacetime interval. Note that while the $i\epsilon$ is subtracted from the inertial time difference in Equation \eqref{Wightman d dim}, the (real and complex) analyticity of the stationary worldlines allows us to subtract the $i\epsilon$ from the proper time in \eqref{wightman d dimensions stationary}. In odd dimensions, the branch of the fractional power is as discussed below \eqref{Wightman d dim}; the denominator has the phases $i^{d-2}$ when $s>0$ and $(-i)^{d-2}$ when $s<0$. 

In order to extract information about the state of the field from the stationary response function, it is useful to express \eqref{stationary RF} in a form where the $\epsilon\to0^+$ limit has been taken under the integral. 
Special cases where this has been addressed in specific spacetime dimensions and/or for specific types of stationary motion are given in \cite{DeWitt1979,Takagi:1986kn,Loukoregspatialprofile,Hodgkinson_2012,Good:2020hav}. 
Here, we present a simple method of finding an $i\epsilon$-independent expression for the stationary response function in any spacetime dimension~$d\ge3$. 

As an example, consider $d=4$, in which case the vacuum Wightman function is
\begin{equation} \label{3+1 stationary Wightman}
    \mathcal{W}(s) = \frac{1}{4\pi^2 \Delta \mathsf{x}^2(s-i\epsilon)}.
\end{equation}
To identify the exact form of the small $s$ distributional behaviour of the Wightman function, we expand the reciprocal of the spacetime interval $\big(\Delta x^2(s-i\epsilon)\big)^{-1}$ as a Laurent series in $(s-i\epsilon)$, keeping only the singular terms as $s\rightarrow 0$
\begin{equation} \label{d=4 square interval singularity}
 \frac{1}{\Delta \mathsf{x} ^2(s-i\epsilon)}= -\frac{1}{(s-i\epsilon)^2} + O(1).
\end{equation}
Adding and subtracting this expansion from the Wightman function within the integral in \eqref{stationary RF}, we obtain
\begin{equation}
    \F(E) = \frac{1}{4\pi^2} \intinf ds e^{-iEs} \Bigg(\frac{1}{\Delta \mathsf{x}^2(s-i\epsilon)} + \frac{1}{(s-i\epsilon)^2} -\frac{1}{(s-i\epsilon)^2}     \Bigg).
\end{equation}
By \eqref{d=4 square interval singularity}, the singularities of the first two terms cancel as $s\rightarrow 0$. In addition, these two terms also individually vanish as $s\rightarrow \infty$ since $\frac{1}{|\Delta \mathsf{x}^2(s)|}$ is bounded by $\frac{1}{s^2}$. Therefore, the sum of the first two terms is integrable independently of the $i\epsilon$ regulator and we can take the $\epsilon \rightarrow 0^+$ limit before integrating. The last term in the brackets can be evaluated by contour integration, after which the $\epsilon \rightarrow 0^+$ limit can be taken. The final expression for the stationary response function is therefore 
\begin{align}
    \F(E) &= \frac{1}{4\pi^2} \intinf ds e^{-iEs} \Bigg(\frac{1}{\Delta \mathsf{x}^2(s)} + \frac{1}{s^2}  \Bigg) - \frac{1}{4\pi^2} \intinf ds \frac{e^{-iEs}}{(s-i\epsilon)^2} \notag \\
    &= \frac{1}{2\pi^2} \intoinf ds \cos(Es) \Bigg(\frac{1}{\Delta \mathsf{x}^2(s)} + \frac{1}{s^2}  \Bigg) - \frac{E}{2\pi} \Theta(-E), \label{rfd=4}
\end{align}
where in the last equality, we have used the fact that $\Delta \mathsf{x}^2(s)$ is even in $s$ to halve the domain of integration and replace the exponential with a cosine. The second term is the response of an inertial detector and the remaining part is the correction due to the non-zero acceleration and torsion. This expression agrees with those found in \cite{Bierman2020, Hodgkinson_2012, Loukoregspatialprofile}.

This method is applicable in Minkowski spacetime of any dimension $d$. In even dimensions $d$, the reciprocal of the spacetime interval contains $(d-2)/2$ divergent terms as $s\rightarrow 0$, while in odd dimensions, it contains $(d-1)/2$ divergent terms as $s\rightarrow 0$. All of these terms need to be subtracted from the Wightman function and the method proceeds in the same way as in $d=4$ dimensions. For general $d$, we give the expressions in Appendix \ref{RF derivation}.

In the main text, we focus on $d=4$, as given above, and $d=3$. Using the expressions obtained in Appendix \ref{RF derivation}, we find the stationary response function in $d=3$ dimensions to be
\begin{equation} \label{2+1 stationary RF}
    \F(E) = \frac{1}{2\pi} \int_0^{\infty} ds \sin(Es) \bigg( \frac{1}{s} - \frac{1}{\sqrt{-\Delta x^2 (s)}} \bigg)  + \frac{1}{2} \Theta(-E),
\end{equation}
which agrees with that found in \cite{Bierman2020,Hodgkinson_2012, Loukoregspatialprofile}.

\section{Temperature asymptotics} \label{Temp asymptotics}

\subsection{Preamble: Effective detailed balance temperature}

Given the stationary response function $\F(E)$~\eqref{stationary RF}, 
we define the frequency-dependent temperature $T(E)$ by 
\begin{equation} \label{effective temp}
    T(E) = \frac{E}{\log \! \left( \frac{\F(-E)}{\F(E)} \right)}. 
\end{equation}
We refer to $T(E)$ as the detailed balance temperature. 
We review here briefly the motivation for this definition. 

For uniform linear acceleration of proper acceleration~$a$, 
we have $T(E)= T_U := a/(2\pi)$, which is independent of~$E$. 
This is the Unruh effect \cite{Unruh1976,Takagi:1986kn,Crispino:2007eb}: 
the detector's response satisfies $\F(-E) = e^{E/T_U} \F(E)$, which is the detailed balance condition between the detector's excitation and de-excitation rates in a Gibbs ensemble of temperature $T_U$ \cite{Einstein,terhaar-book,Fredenhagen:1986jg,Kubo:1957mj,Martin:1959jp}. What is behind this outcome is that the Minkowski vacuum can be expressed as a genuine thermal ensemble of excitations over the vacuum defined by the boost Killing vector that generates the accelerated motion. An essential part of this description is that the Killing horizon of the boost Killing vector divides the spacetime into four quadrants, and the thermality in one Rindler quadrant where the Killing vector is timelike arises from tracing out the field degrees of freedom in the opposite, causally disconnected Rindler quadrant \cite{Unruh1976,Takagi:1986kn,Crispino:2007eb}. 

By contrast, for other types of non-inertial stationary motion,
$T(E)$ depends on~$E$ and there appears to be no known way to associate $T(E)$ with an  underlying Gibbs ensemble. 
For example, in the case of uniform circular motion, 
the Killing vector generating the motion changes from timelike to spacelike at the speed-of-light surface, 
and one might therefore expect the speed-of-light surface to play a role analogous to that of the Rindler horizon. However, while the detector's response in circular motion is nontrivial, the spacelike character of the Killing vector everywhere outside the speed-of-light surface creates technical obstacles to attempts to define a ``rotating vacuum'' on which the Minkowski vacuum could be interpreted as excitations \cite{Denardo:1978dj,Letaw:1979wy,Davies:1996ks,Unruh:2022gso}.

We emphasise that as we only address the transitions in the detector without observing the transitions in the field, the absence of a ``vacuuum'' adapted to the detector's motion plays no role in the analysis, nor does the fact that the Killing vector generating the trajectory becomes spacelike far from the trajectory and does therefore not provide a global notion of time evolution in Minkowski. These issues will however play a role if one wishes to observe the ``particles'' that the interaction with the detector emits into the field~\cite{UnruhRindlerparticle}. 

All of the above being said, the utility of the detailed balance temperature $T(E)$ \eqref{effective temp}, even when $E$-dependent, is that it provides a useful quantifier of the detector's response to acceleration at a given energy scale. One example of this is that in the Born-Markov approximation, the late-time asymptotic state of the detector is \cite{Juarez-Aubry:2019gjw}
\begin{equation}
    \rho(E) = \frac{1}{1+e^{-E/T(E)}} 
    \begin{pmatrix}
        1 & 0 \\
        0 & e^{-E/T(E)}
    \end{pmatrix} . 
\end{equation}

In the rest of Section~\ref{Temp asymptotics}, we investigate the large gap, small gap and ultrarelativistic limits of $T(E)$ \eqref{effective temp} for drifted Rindler motion in 3+1 and 2+1 dimensions, comparing the outcomes with those of circular motion~\cite{Bierman2020}.

\subsection{Drifted Rindler motion in 3+1 dimensions} \label{contour section}
Consider drifted Rindler motion in 3+1 dimensions. From \eqref{DR worldline} and \eqref{rfd=4}, the response function can be split into the inertial contribution $\F^\inert(E)$ and the non-inertial correction $\F^\corr(E)$ as 
\begin{subequations}
\begin{align} 
\F(E) &= \F^\inert(E) + \F^\corr(E), 
\label{RF DR split}
\\
\F^\inert(E) &= -\frac{E}{2\pi}\Theta(-E), \label{RF DR in}\\
\mathcal{F}^{\text{corr}}(E) &= \frac{1}{4\pi^2 \gamma  R} \int_0^\infty dz \cos(\tfrac{2ER}{\gamma }z) \Bigg(\frac{\gamma^2}{z^2}-\frac{1}{\sinh^2z-v^2z^2} \Bigg), \label{RF DR corr}
\end{align}
\end{subequations}
where in \eqref{RF DR corr} we have changed variables to $z= \frac{\gamma}{2R}s$. In order to evaluate the large gap and ultrarelativistic limits, it is useful to re-express \eqref{RF DR corr} as an integral along a contour in the complex $z$ plane. 
Following Appendix C of \cite{hodgkinsonstatic}, we 
first extend the integral in \eqref{RF DR corr} to the full real axis, using the evenness of the integrand. We then replace $\cos(\tfrac{2ER}{\gamma }z)$ by $\exp(i\tfrac{2|E|R}{\gamma }z)$. Next, we deform the contour to a new contour $C$ that bypasses $z=0$ in the upper half-plane, say, along a small half-circle. The contribution from the first term in the parentheses then vanishes, as seen by closing $C$ 
in the upper half-plane, and what remains is 
\begin{equation} \label{RF DR contourcorr}
    \mathcal{F}^{\text{corr}}(E) = -\frac{1}{8\pi^2 \gamma R} \int_C dz \, \frac{\exp(i\tfrac{2|E|R}{\gamma }z)}{\sinh^2z-v^2z^2}.
\end{equation}
Closing $C$ in \eqref{RF DR contourcorr} in the upper half-plane shows that 
\eqref{RF DR contourcorr} equals the sum of residues at the poles in the upper half-plane. 

An analysis of the zeroes of the function $g(z) = \sinh^2\! z - v^2 z^2$ in the upper half-plane is given in Appendix \ref{sing structure SECTION}. We summarise the outcomes here. 
For a given speed $v$, there are finitely many purely imaginary zeroes, which we write as $z_k=i\alpha_k$ with 
$k=0,1,\dots, N$ and $0<\alpha_0<\alpha_1<\cdots <\alpha_N$;
these zeroes are simple, except that $\alpha_N$ is a double zero 
when $v^2$ is a local maximum value of $\frac{\sin^2 \! \alpha}{\alpha^2}$. 
$\alpha_0$ is in the interval $0<\alpha_0<\pi$, 
and we may parametrise $v$ in terms of $\alpha_0$ as
\begin{equation} \label{v and alpha0}
    v = \frac{\sin\alpha_0}{\alpha_0} . 
\end{equation}
In addition, there is a countable infinity of simple zeroes that have both non-zero real and imaginary parts. We write these zeroes as $z_k = i(\alpha_k \pm i\beta_k)$, where $k=N+1,N+2,\dots$, $\alpha_N<\alpha_{N+1}<\cdots$, and $\beta_k>0$. For each $\alpha_k$ with $k>N$, 
there are hence two zeroes, with real parts of equal magnitude but opposite sign. 

Now, applying the residue theorem, we find that when 
$v^2$ is not a local maximum value of $\frac{\sin^2 \! \alpha}{\alpha^2}$, 
\eqref{RF DR contourcorr} is equal to 
\begin{subequations}
\label{eq:3+1Fcorr-sum}
    \begin{align}
    \F^\corr(E) &= \mathcal{F}_{\im}^{\corr}(E) + \mathcal{F}_{\comp}^{\corr}(E), \\
    \F_\im^\corr(E) &= \frac{1}{8\pi R} \sum_{k=0}^{N} \frac{\sqrt{\alpha_k^2-\sin^2\alpha_k}  }{\sin\alpha_k(\sin\alpha_k-\alpha_k\cos\alpha_k)} e^{-\tfrac{2|E|R}{\gamma}\alpha_k}, \label{3+1 im RF} \\
    \F_\comp^\corr(E) &= \frac{1}{8\pi\gamma v R} \sum_{k=N+1}^\infty \frac{\exp\Big( -\tfrac{2|E|R}{\gamma}(\alpha_k+i\beta_k) \Big)}{(\alpha_k+i\beta_k)(1-\alpha_k\cot\alpha_k+i\beta_k\tan\alpha_k)} + (\beta_k \rightarrow -\beta_k), \label{3+1 comp RF}
\end{align}
\end{subequations}
where $\F_\im^\corr(E)$ and $\F_\comp^\corr(E)$ are the respective contributions from the $N+1$ purely imaginary poles and from the countably many poles with nonvanishing real and imaginary parts. 
When $v^2$ is a local maximum value of $\frac{\sin^2 \! \alpha}{\alpha^2}$, the $k=N$ term in \eqref{3+1 im RF} is replaced by 
\begin{align}
- \frac{1}{\pi R} \left(\frac{|E|R}{2} + \frac{\sqrt{1 + \alpha_N^2}}{3\alpha_N^2}\right)
e^{-\tfrac{2|E|R}{\gamma}\alpha_N} . 
\end{align}

\subsubsection{Large gap limit}
Consider the large gap limit $|E|\rightarrow \infty$ while keeping $v$ and $R$ fixed. To calculate the temperature in this limit, we follow the method outlined in Section 3.2 of \cite{Bierman2020}. Due to the exponential term appearing in \eqref{3+1 im RF} and \eqref{3+1 comp RF}, the pole with the smallest magnitude $z=i\alpha_0$ dominates in the large gap limit. The detailed balance temperature \eqref{effective temp} in this limit is the reciprocal of the coefficient in the exponent, 
\begin{equation} \label{large E temp 3+1}
    T(E) = \frac{\gamma}{2\alpha_0 R}.
\end{equation}

Recall that the linear acceleration Unruh effect prediction for the temperature is $T_\lin=\frac{a}{2\pi}$, where the proper acceleration $a$ for drifted Rindler motion is given in~\eqref{large E temp 3+1}, 
and it depends on both $v$ and~$R$. 
For the ratio of $T(E)$ and $T_\lin$, we find 
\begin{equation}
\label{eq:TDR/Tlin-ratio}
    \frac{T(E)}{T_\lin} = \pi \frac{\sqrt{\alpha_0^2-\sin^2\alpha_0}}{\alpha_0^2} \in \left(1,\frac{\pi}{\sqrt{3}}\right). 
\end{equation}
This ratio depends only on~$v$, approaching the lower bound $1$ as $v\to 0$ ($\alpha_0 \to \pi$) 
and 
the upper bound $\frac{\pi}{\sqrt{3}}$ as $v \to 1$ ($\alpha_0\to0$), 
and being monotonic in $v$ in between. 
In particular, the ratio is always greater than~$1$. The drift speed $v$ thus has a heating effect relative to the linear acceleration Unruh temperature, but by a factor never exceeding 
$\frac{\pi}{\sqrt{3}} \approx 1.81$. 

It is instructive to compare the large gap ratio \eqref{eq:TDR/Tlin-ratio} to the similar large gap ratio for circular motion~\eqref{CM worldlime}. 
This ratio was found in \cite{Bierman2020}, and it satisfies 
\begin{equation}
    \frac{T_\CM(E)}{T_\lin} \in \left( \frac{\pi}{\sqrt{3}}, \infty \right),
\end{equation}
where the lower and upper bounds correspond respectively to $v\rightarrow 1$ and $v\rightarrow 0$, and $v$ is the circular motion orbital speed in~\eqref{CM worldlime}. 
We see that the critical value $\frac{\pi}{\sqrt3}$ emerges as the demarcation point between the ratios obtained in the two families of motions: in the drifted Rindler motion, the critical value is approached from below as $v\to1$, and in circular motion it is approached from above as $v\to1$. 
This result is consistent with the geometric connection between drifted Rindler motion and circular motion discussed in Section~\ref{DR CM Geometry}.

\subsubsection{Small gap limit}
Consider next the small gap limit $E\rightarrow 0$ while keeping $v$ and $R$ fixed. 
We show in Appendix \ref{APP small gap RF} that the small $E$ expansion of the stationary response function to first order in $E$ is
\begin{equation}
    \F(E) = \frac{1}{4\pi^2\gamma R} \int_0^\infty dz \left( \frac{\gamma^2}{z^2} - \frac{1}{\sinh^2z-v^2z^2} \right)-\frac{E}{4\pi} +O(E^2).
\end{equation}
It follows that the detailed balance temperature \eqref{effective temp} has the small $E$ expansion 
\begin{equation}
    T(E) =  J(v) T_\lin + O(E) , 
\end{equation}
where $T_\lin = \frac{a}{2\pi}$ is the linear acceleration prediction 
and 
\begin{equation} \label{J(v)}
    J(v) = \int_0^\infty dz \left( \frac{1}{\gamma z^2} - \frac{1}{\gamma^3(\sinh^2z-v^2z^2)} \right).
\end{equation}
We show in Appendix \ref{J(v) limits} that $J(v)$ decreases monotonically from $1$ when $v=0$ to $\frac{\pi}{2\sqrt{3}} \approx 0.91$ when $v\to 1$. 
The drift speed $v$ thus has a mild cooling effect, by less than 10\%, relative to the linear acceleration Unruh temperature.

\subsubsection{Ultrarelativistic limit}
\label{sec:3+1ultrarel}

Consider finally the ultrarelativistic limit $v\rightarrow 1$ with $R$ fixed. 
In terms of $\alpha_0$, this is the limit $\alpha_0\to0$, and $v$, $\gamma$ and $a$ have the expansions 
\begin{subequations} \label{small a0 expansions}
  \begin{align}
    v &= \Bigg(1 - \frac{\alpha_0^2}{6}\Bigg)\Bigl(1 + O\bigl(\alpha_0^2\bigr)\Bigr), \\
    \gamma &= \frac{\sqrt{3}}{\alpha_0}\Bigl(1 + O\bigl(\alpha_0^2\bigr)\Bigr), \\ 
    a &= \frac{\gamma^2}{R} = \frac{3}{\alpha_0^2R} \Bigl(1 + O\bigl(\alpha_0^2\bigr)\Bigr).
\end{align}  
\end{subequations}
It hence suffices to consider $\alpha_0$ so small that \eqref{3+1 im RF} and \eqref{3+1 comp RF}
hold with $N=0$. In $\F_\im^\corr(E)$ \eqref{3+1 im RF}, the only term is $k=0$, giving 
\begin{align}
\F_\im^\corr(E) 
&= 
\frac{1}{8\pi R} \frac{\sqrt{\alpha_0^2-\sin^2\alpha_0}  }{\sin\alpha_0(\sin\alpha_0-\alpha_0\cos\alpha_0)} e^{-\tfrac{2|E|R}{\gamma}\alpha_0}
\notag\\[1ex]
&= 
\frac{\sqrt{3} \exp \! \left(-\frac{2\alpha_0|E|R}{\gamma}\right)}{8\pi R\alpha_0^2} \Bigl(1 + O\bigl(\alpha_0^2\bigr)\Bigr) . 
\label{eq:Fim3+1exp}
\end{align}
In $\F_\comp^\corr(E)$ \eqref{3+1 comp RF}, the analysis in Appendix 
\ref{sing structure SECTION} shows that $\pi k \le \alpha_k$ and $c_2 k \le \beta_k \tan\alpha_k$ for all $k=1,2,\ldots$, where $c_2$ is a purely numerical positive constant.
In each summand in \eqref{3+1 comp RF}, the absolute value of the denominator is hence bounded below by $\pi c_2 k^2$. It then follows by a dominated convergence argument that 
\begin{align}
\F_\comp^\corr(E) 
&= 
O \Biggl( \alpha_0 \exp \! \left(-\frac{2\pi|E|R}{\gamma}\right) \Biggr), 
\label{eq:Fcomp3+1exp}
\end{align}
uniformly in~$E$. 
Combining 
\eqref{RF DR split}, 
\eqref{eq:3+1Fcorr-sum}, 
\eqref{eq:Fim3+1exp}
and 
\eqref{eq:Fcomp3+1exp}, we find 
\begin{align} \label{RF v 1}
    \F(E) &= -\frac{E}{2\pi}\Theta(-E) + \frac{\sqrt{3} \exp \! \left(-\frac{2\alpha_0|E|R}{\gamma}\right)}{8\pi R\alpha_0^2} \Bigl(1 + O\bigl(\alpha_0^2\bigr)\Bigr) 
    + O \Biggl( \alpha_0 \exp \! \left(-\frac{2\pi|E|R}{\gamma}\right) \Biggr)
    \notag
    \\
 &= -\frac{E}{2\pi}\Theta(-E) 
+ \frac{\sqrt{3} \exp \biggl(-2 \sqrt3 \frac{|E|}{a} \Bigl(1 + O\bigl(\alpha_0^2\bigr)\Bigr)\biggr)}{8\pi R\alpha_0^2} \Bigl(1 + O\bigl(\alpha_0^2\bigr)\Bigr) 
    + O \Biggl( \alpha_0 \exp \! \left(-\frac{2 \sqrt3 \, \pi}{\alpha_0} \frac{|E|}{a}\right) \Biggr) , 
\end{align}
where in the last equality we have used~\eqref{small a0 expansions}. 
Note that the error terms in the last expression in \eqref{RF v 1} are uniform in $|E|/a$. 

For the detailed balance temperature, \eqref{RF v 1} gives the $v\to1$ limit 
\begin{equation} \label{3+1 ultrarel T}
    T(E) = \frac{|E|}{\log \! \Bigg( 1 + 4\sqrt{3} \frac{|E|}{a} \exp \! \bigg(2\sqrt{3} \frac{|E|}{a} \bigg) \Bigg)}.
\end{equation}
Formula \eqref{3+1 ultrarel T} 
is the detailed balance temperature in the ultrarelativistic limit of circular motion \cite{Bierman2020,Unruh:1998gq}, 
and it is also the detailed balance temperature in parator motion~\cite{Good:2020hav}. 
This is another consequence of the geometric connection between the circular and drifted Rindler motions through the parator motion. 

\subsection{Drifted Rindler motion in 2+1 dimensions}
We now turn to drifted Rindler motion in 2+1 dimensions. Using \eqref{CM worldlime} and \eqref{2+1 stationary RF}, the response function can be split into the inertial contribution $\F^\inert(E)$ and the non-inertial correction $\F^\corr(E)$ as
\begin{subequations}
    \begin{align}
        \F(E) &= \F^\inert(E) + \F^\corr(E), \\
        \F^\inert(E) &= \frac{1}{2}\Theta(-E), \\
        \F^\corr(E) &= \frac{1}{2\pi \gamma } \int_0^\infty dz \sin(\tfrac{2ER}{\gamma }z) \Bigg(\frac{\gamma }{z}-\frac{1}{\sqrt{\sinh^2z-v^2z^2}} \Bigg).
        \label{eq:Fcorr2+1}
    \end{align}
\end{subequations}
It will also be useful to split the response function into an even and an odd part as
\begin{subequations}
\label{eq:F2+1evenplusodd}
    \begin{align} \label{RF odd even d=3}
        \F(E) &= \F^\text{even}(E) + \F^\text{odd}(E), \\
        \F^\text{even}(E) &= \frac{1}{4}, \\
        \F^\text{odd}(E) &= -\frac{1}{2\pi \gamma} \int_0^{\infty} dz \frac{\sin \! \left(\frac{2ER}{\gamma}z\right)}{\sqrt{\sinh^2z-v^2z^2}}. \label{RFodd}
\end{align}
\end{subequations}
We note that $\F^\corr(E)$ \eqref{eq:Fcorr2+1} can be written as a sum over contour integrals in the complex plane, encircling the branch points at the zeroes of the square root in the denominator, as done for circular motion in Section 4.1 of \cite{Bierman2020}. 
The convergence of the sums is however weaker than in the corresponding sums over residues in \eqref{3+1 comp RF}, and we shall not be using this sum here.

\subsubsection{Large gap limit}

Consider the large gap limit $|E| \to\infty$ while keeping $v$ and $R$ fixed. 
Starting from $\F^\corr(E)$ \eqref{eq:Fcorr2+1} and deforming the integration contour in the complex plane as in Sections 4.1 and 4.2 of \cite{Bierman2020}, we find that the dominant contribution to $\F^\corr(E)$ comes from the vicinity of the branch point $z=i\alpha_0$, and this contribution has the exponential factor $\exp \! \left(-\tfrac{2\alpha_0|E|R}{\gamma}\right)$. It follows as in $3+1$ dimensions that the detailed balance temperature is given by~\eqref{large E temp 3+1}.

\subsubsection{Small gap limit} \label{SEC DR small gap 2+1}
Consider next the small gap limit $E\rightarrow 0$ with $v$ and $R$ fixed. 
We begin by writing $\F^\odd(E)$ \eqref{RFodd} as
\begin{equation}
    \F^\odd(E) = -\frac{ER}{\pi\gamma^2} \int_0^\infty dz \frac{\sin \! \left(\frac{2ER}{\gamma}z\right)}{\frac{2ER}{\gamma}z} \frac{z}{\sqrt{\sinh^2 \! z-v^2z^2}}.
\label{eq:Fodd2+1-sinc}
\end{equation}
As $z(\sinh^2 \! z-v^2z^2)^{-\frac{1}{2}}$ is an integrable function with an exponential fall-off at $z\to\infty$, a dominated convergence argument shows that 
the small $E$ asymptotic expansion of $\F^\odd(E)$ is obtained by expanding $\frac{\sin x}{x}$ in the integrand in \eqref{eq:Fodd2+1-sinc} in the Maclaurin series. 
From \eqref{eq:F2+1evenplusodd}, we then have 
\begin{equation}
    \F(E) =\frac{1}{4} -\frac{ER}{\pi\gamma^2}\int_0^\infty dz \frac{z}{\sqrt{\sinh^2 \! z-v^2z^2}} + O(E^3).
\label{eq:F2+1smallgap}
\end{equation}

From \eqref{eq:F2+1smallgap}, the detailed balance temperature is 
\begin{equation} \label{d=3 small gap temp}
    T(E) = \frac{T_\lin}{ K(v)} + O(E^2) , 
\end{equation}
where
\begin{equation} \label{K(v)}
    K(v) = \frac{4}{\pi^2} \int_0^\infty dz \frac{z}{\sqrt{\sinh^2 \! z-v^2z^2}}.
\end{equation}
In Appendix \ref{J(v) limits}, we show that $K(v)$ increases monotonically from $1$ to $\infty$ as $v$ increases from $0$ to~$1$. Therefore, from \eqref{d=3 small gap temp}, we see that the drift speed $v$ cools the small gap temperature relative to the linear Unruh temperature, by a factor that increases without bound as $v$ approaches~$1$. 

In comparison, the circular motion detailed balance temperature in $2+1$ dimensions vanishes linearly in $E$ in the small gap limit~\cite{Bierman2020}, as 
\begin{equation} \label{small gap circular}
    T_\CM(E) = \frac{|E|}{\displaystyle{\log \! \left( \frac{\gamma_\CM+1}{\gamma_\CM-1}\right)}} + O(E^2).
\end{equation}
We shall return to this comparison in Section~\ref{Exceptionality section}.

\subsubsection{Ultrarelativistic limit with fixed $|E|/a$} \label{d=3 v->1 section}

Consider the ultrarelativistic $v\rightarrow 1$ limit with fixed $|E|/a$. We show in Appendix \ref{Append 2+1 ultrarel lim} that in the $v\rightarrow 1$ limit with fixed $|E|/a$, $\F^\odd(E)$ has the limiting behaviour
\begin{equation} \label{RF odd v->1}
    \F^\text{odd}(E) \rightarrow -\frac{1}{2\pi} G \bigl(2\sqrt{3}E/a\bigr) , 
\end{equation}
where 
\begin{equation} \label{G(q)}
    G(q) := \intoinf dx \frac{\sin(qx)}{x\sqrt{1+x^2}}.
\end{equation}
As a result, the detailed balance temperature in this limit is 
\begin{equation} \label{d=3 v->1 temp}
   T(E) = \frac{|E|}{\displaystyle \log \! \Bigg( \frac{1+\frac{2}{\pi}G\big(2\sqrt{3}|E|/a\big)}{1-\frac{2}{\pi}G\big(2\sqrt{3}|E|/a\big)}  \Bigg)}.
\end{equation}

The detailed balance temperature \eqref{d=3 v->1 temp} agrees with the $2+1$ circular motion detailed balance temperature in the ultrarelativistic limit with fixed $|E|/a$~\cite{Bierman2020}. 
This temperature is precisely the detailed balance temperature in parator motion in $2+1$ dimensions: from \eqref{PA worldline}, 
parator motion has $\Delta\mathsf{x}^2(s) = -s^2-\frac{a^2}{12}s^4$, 
and substituting this into \eqref{RF odd even d=3} gives 
\begin{align}
    \F_\PA (E) &= \frac{1}{4} - \frac{1}{2\pi} \intoinf ds \frac{\sin(Es)}{s\sqrt{1+\frac{a^2}{12}s^2}} \notag \\
    &=  \frac{1}{4} - \frac{1}{2\pi} G\bigl(2\sqrt{3} E/a\bigr).
\end{align}
The geometric connection between the circular and drifted Rindler motions through parator motion hence also extends to the detailed balance temperature in $2+1$ dimensions, despite the differing small gap behaviour. 
We shall address the small gap behaviour in the different motions in more detail in Section~\ref{Exceptionality section}.

\section{Exceptionality of 2+1 circular and parator motion in the small gap limit} \label{Exceptionality section}

In 2+1 dimensions, it was shown in \cite{Bierman2020} that the circular motion temperature vanishes in the small gap limit, whereas we showed in Section \ref{SEC DR small gap 2+1} that the drifted Rindler motion temperature does not vanish in this limit. By contrast, in 3+1 dimensions, neither the circular motion temperature nor the drifted Rindler temperature vanishes in the small gap limit. A natural question to ask is then: for which types of stationary motion in which dimensions does the temperature vanish in the small gap limit?

The central observation is that the detailed balance temperature \eqref{effective temp} has a finite positive limit $T_0$ as $E\to 0$ if and only if the response function satisfies $\F(-E)/\F(E)= 1+\frac{1}{T_0}E + o(E)$ as $E\to 0$. If $\F(-E)/\F(E)$ has different positive limits as $E\to0^+$ and $E\to0^-$, the small gap temperature vanishes. 
If $\F(-E)/\F(E) \to 1$ as $E\to 0$ but the next-to-leading order term decays less rapidly than $O(E)$, then the small gap temperature vanishes, whereas if the next-to-leading term decays more rapidly than $O(E)$, then the small gap temperature diverges. 

In this section, we calculate the small gap expansion of the stationary response function for all types of stationary motion in all dimensions, 2+1 and higher. We shall see that in all except two cases this expansion is of the form
\begin{equation} \label{RF small E}
    \F(E) = \alpha - (\beta+\eta \sgn E) E + o(E),
\end{equation}
where $\alpha$ and $\beta$ are positive constants and $\eta$ is a real-valued constant. 
From \eqref{RF small E} it follows that 
$\F(-E)/\F(E) = 1 + \frac{2\beta}{\alpha}E+o(E)$ as $E\to 0$ and hence the resulting small gap temperature is
\begin{equation} \label{T small E}
    T(E) = \frac{\alpha}{2\beta}\bigl( 1+ o(1) \bigr).
\end{equation}
Note that \eqref{T small E} does not involve the constant~$\eta$.  

The two exceptions where \eqref{RF small E} does not hold are circular and parator motion in 2+1 dimensions. This phenomenon can be traced back to the weak decay of the Wightman function for these types of motion in 2+1 dimensions. The weak decay gives the response function a small gap behaviour that makes the small gap temperature $O(E)$ for circular motion and $O(1/\log |E|)$ for parator motion. This shows that the small gap temperature vanishes only in 2+1 dimensions and therein only for circular and parator motion.

\subsection{2+1 dimensions}
In 2+1 dimensions, there are five types of stationary motion, and we have compiled them in Table \ref{TABLE}. We exclude inertial motion as the temperature is identically zero. The stationary response function \eqref{2+1 stationary RF} can be written as 
\begin{equation} \label{2+1 stationary RF 3}
    \F(E) = \frac{1}{4} - \frac{E}{2\pi}\int_0^\infty ds \frac{\sin(Es)}{Es} \frac{s}{\sqrt{-\Delta \mathsf{x}^2(s)}}.
\end{equation}
We consider the small $E$ expansion of \eqref{2+1 stationary RF 3}.

If the expression
\begin{equation} \label{2+1 integrable?}
    \frac{s}{\sqrt{-\Delta\mathsf{x}^2(s)}}
\end{equation}
is integrable, then we may use dominated convergence in \eqref{2+1 stationary RF 3} to write
\begin{equation} \label{2+1 linear RF}
    \F(E) = \frac{1}{4} - \frac{E}{2\pi}\int_0^\infty ds \frac{s}{\sqrt{-\Delta \mathsf{x}^2 (s)}} + o(E).
\end{equation}
This is of the form of \eqref{RF small E} and so the corresponding small gap temperature is a non-zero constant given by \eqref{T small E}.

Whether \eqref{2+1 integrable?} is integrable depends on the asymptotic behaviour of the type of motion under consideration. For small $s$, \eqref{2+1 integrable?} is $O(1)$, since $\Delta\mathsf{x}^2(s) = -s^2 + O(s^4)$ as $s\to 0$ for all types of stationary motion, the acceleration affecting only order $O(s^4)$. However, for large $s$, if the decay of \eqref{2+1 integrable?} is not sufficiently strong, then \eqref{2+1 integrable?} may not be integrable. 

The large $s$ asymptotic behaviour of $\Delta \mathsf{x}^2(s)$ for each type of stationary motion is compiled in Table \ref{TABLE}. The types of motion for which \eqref{2+1 integrable?} is integrable are uniform linear acceleration (Rindler motion) and drifted Rindler motion. In these cases, \eqref{2+1 integrable?} is exponentially decaying and therefore, by dominated convergence, we arrive at \eqref{2+1 linear RF}. Hence for these types of motion, the small gap temperature is finite and non-zero.

The types of motion (aside from inertial motion) for which \eqref{2+1 integrable?} is not integrable are circular motion and parator motion. For circular motion, \eqref{2+1 integrable?} is $O(1)$ as $s\to \infty$, and for parator motion, \eqref{2+1 integrable?} decays only as $O(1/s)$ as $s\to \infty$. Therefore, we cannot simply use dominated convergence in \eqref{2+1 stationary RF 3}. We must instead consider these two cases individually.

Firstly, consider circular motion, for which the large $s$ asymptotic behaviour of the spacetime interval is $\Delta\mathsf{x}^2(s) = -\gamma^2s^2 + O(1)$. Following the method in Appendix B of \cite{Bierman2020}, 
we add and subtract the large $s$ behaviour of the integrand as follows
\begin{equation} \label{2+1 CM RF E->0}
    \F(E) = \frac{\gamma-\sgn E}{4\gamma} - \frac{E}{2\pi} \int_0^\infty ds \frac{\sin(Es)}{Es}\left( \frac{s}{\sqrt{-\Delta \mathsf{x}^2(s)}} - \frac{1}{\gamma}   \right),
\end{equation}
where we have used 
\begin{equation}
    \int_0^\infty ds \frac{\sin(Es)}{s} =  \frac{\pi}{2} \sgn E.
\end{equation}
The expression in parentheses in \eqref{2+1 CM RF E->0} is of order $O(s^{-2})$ as $s\to \infty$ and hence, by dominated convergence,
\begin{equation} \label{RF CM small E 2+1}
    \F(E) = \frac{\gamma-\sgn E}{4\gamma} + O(E).
\end{equation}
Since the zeroth order term in \eqref{RF CM small E 2+1} depends on the sign of $E$, the temperature vanishes in the small gap limit. The temperature to leading order in $E$ is given by \eqref{small gap circular},
\begin{equation} 
    T(E) = \frac{|E|}{\displaystyle{\log\left( \frac{\gamma_\CM+1}{\gamma_\CM-1}\right)}} + O(E^2).
\end{equation}

Secondly, consider parator motion. The spacetime interval is $\Delta \mathsf{x}^2(s) = -s^2-\frac{a^2}{12} s^4$ and thus \eqref{2+1 stationary RF 3} becomes
\begin{equation} \label{RF E->0 PA}
    \F(E) = \frac{1}{4} - \frac{1}{2\pi}G\!\left( 2\sqrt{3} \frac{E}{a} \right),
\end{equation}
where $G(q)$ is given by \eqref{G(q)}. The leading small argument behaviour of $G(q)$ was calculated in Appendix D of \cite{Bierman2020} to be $G(q) = -q\log|q| + O(q)$.  With this result, \eqref{RF E->0 PA} becomes
\begin{equation} 
    \F(E) = \frac{1}{4} + \frac{\sqrt{3}}{\pi} \frac{E}{a}\log\!\left(\frac{|E|}{a}\right) + O(E)
\end{equation}
as $E\to 0$. The next-to-leading order term goes to zero less rapidly than $O(E)$, and the temperature vanishes in the small gap limit. The temperature to leading order in $E$ is given by
\begin{equation} \label{parator small gap temp}
    T(E) = \frac{\pi a}{8\sqrt{3}\log \bigl( a/|E| \bigr)} \big( 1 + o(1) \big).
\end{equation}
The $1/\log\bigl( a/|E| \bigr)$ suppression in \eqref{parator small gap temp} 
is reminiscent of the $1/\log\gamma$ suppression in the drifted Rindler small gap temperature \eqref{d=3 small gap temp} in the ultrarelativistic limit $v\to1$, as seen from~\eqref{K(v) ref}. 

The vanishing of the small gap temperatures for circular and parator motion is therefore a result of the weak decay of their respective Wightman functions in 2+1 dimensions.

\begin{table}
    \centering
    \begin{center}
        \begin{tabular}{ |c|c|c| } 
        \hline
        Type of motion & $\Delta \mathsf{x}^2(s)$ & $s\to \infty$ \\
        \hline
        Inertial & $-s^2$ & $O(s^2)$ \\ 
        Rindler & $-\frac{4}{a^2}\sinh^2\!\left(\tfrac{as}{2}\right)$ & $O(e^{as})$ \\ 
        Parator & $-s^2-\frac{a^2}{12}s^4$ & $O(s^4)$ \\ 
        Circular & $\gamma^2 s^2 + 4R^2\sin^2\!\left(\frac{\gamma v s}{2R}  \right)$ & $O(s^2)$ \\
        Drifted Rindler & $\gamma^2 v^2 s^2 - 4R^2 \sinh^2\!\left( \frac{\gamma s}{2R} \right)$ & $O(e^{\gamma s/R})$ \\
        \hdashline
        Loxodromic & $4R^2\sin^2\!\left( \frac{\gamma v s}{2R} \right) - \frac{4}{a^2}\sinh^2\!\left( \frac{\gamma a s}{2}\right)$  & $O(e^{\gamma a s})$ \\
        \hline
        \end{tabular}
    \end{center}
    \caption{A list of all types of stationary motion in 2+1 and 3+1 dimensions and the $s\to\infty$ asymptotic behaviour of their spacetime intervals $\Delta \mathsf{x}^2(s)$. Loxodromic motion only occurs in 3+1 dimensions.}
    \label{TABLE}
\end{table}

\subsection{3+1 dimensions}
In 3+1 dimensions, there is one more type of stationary motion: loxodromic motion. It is the orbit of a boost with proper acceleration $a$ and a rotation with orbital speed $v$. The worldline can be found in \cite{Good:2020hav}. In Table \ref{TABLE}, we present just the spacetime interval and its large proper time asymptotic behaviour.

In 3+1 dimensions, the response function is given by \eqref{rfd=4} 
\begin{equation} \label{RF 3+1 E->0}
    \F(E) = \frac{1}{2\pi^2}\int_0^\infty ds \cos(Es) \left( \frac{1}{\Delta\mathsf{x}^2(s)} + \frac{1}{s^2} \right)- \frac{E}{2\pi}\Theta(-E).
\end{equation}
We consider the small $E$ expansion of \eqref{RF 3+1 E->0}. For small $s$, the expression in parentheses is $O(1)$ since $\Delta\mathsf{x}^2(s) = -s^2 + O(s^4)$ as $s\to 0$ for all types of stationary motion. Furthermore,
$ \left|\frac{1}{\Delta\mathsf{x}^2(s)}\right| \leq \frac{1}{s^2}$ and thus the decay of the expression in parentheses at $s\to \infty$ is at least $O(1/s^2)$. Therefore, this expression is integrable and we can split \eqref{RF 3+1 E->0} as
\begin{subequations}
    \begin{align} 
    \F(E) &=  \F(0) + \frac{E^2}{2\pi^2}\int_0^\infty ds \frac{(\cos(Es)-1)}{(Es)^2}\left(\frac{s^2}{\Delta\mathsf{x}^2(s)} + 1 \right) - \frac{E}{2\pi}\Theta(-E), \label{RF 3+1 E->0 2}\\ 
    \F(0) &= \frac{1}{2\pi^2}\int_0^\infty ds \left( \frac{1}{\Delta\mathsf{x}^2(s)} + \frac{1}{s^2} \right),
\end{align}
\end{subequations}
where we have multiplied and divided by $(Es)^2$ in the second integral in \eqref{RF 3+1 E->0 2}. We can further simplify \eqref{RF 3+1 E->0 2} as follows
\begin{align}
   \F(E) &= \F(0) + \frac{E^2}{2\pi^2} \int_0^\infty ds \frac{(\cos(Es)-1)}{(Es)^2} \left(\frac{s^2}{\Delta\mathsf{x}^2(s)} \right) -\frac{|E|}{4\pi} - \frac{E}{2\pi}\Theta(-E) \notag\\
    &= \F(0) + \frac{E^2}{2\pi^2} \int_0^\infty ds \frac{(\cos(Es)-1)}{(Es)^2} \left(\frac{s^2}{\Delta\mathsf{x}^2(s)} \right) -\frac{E}{4\pi}, \label{RF 3+1 E->0 3}
\end{align}
where in the first equality we have used the standard integral given in \eqref{cos integral}, and in the second equality we have used $|E| + 2E\Theta(-E) = E$.

Consider the integral term in \eqref{RF 3+1 E->0 3}. The function $\big( \cos(Es)-1 \big)/(Es)^2$ is bounded in absolute value by an $E$-independent constant and it has the pointwise limit $-\frac{1}{2}$ as $E\to 0$. If the function
\begin{equation} \label{3+1 integrable??}
    \frac{s^2}{\Delta\mathsf{x}^2(s)}
\end{equation}
is integrable, then by dominated convergence, the second term in \eqref{RF 3+1 E->0 3} is $O(E^2)$ as $E\to 0$. In this case, the small $E$ expansion \eqref{RF 3+1 E->0 3} is of the form \eqref{RF small E} with $\eta =0$. Hence the small gap temperature is finite and non-zero. 

As can be seen from Table \ref{TABLE}, \eqref{3+1 integrable??} is integrable for all types of stationary motion apart from circular motion (and inertial motion). Although, \eqref{3+1 integrable??} is not integrable for circular motion, we show in Appendix \ref{APP small gap RF} that the small gap expansion of the response function is of the form \eqref{RF small E}, but now with $\eta \neq 0$. As noted below \eqref{T small E}, $\eta$ does not contribute to the temperature. Therefore, the small gap temperature is finite and non-zero for all types of stationary motion in 3+1 dimensions.

\subsection{Higher dimensions}
In the previous two subsections, we saw that the small $E$ behaviour of the response function is sensitive to the large proper time asymptotics of the Wightman function. In this subsection, we show that the bound $ \frac{1}{\left|\Delta \mathsf{x}^2(s)\right|}  \geq \frac{1}{s^2}$ means the Wightman function has sufficiently strong decay at $s\to \infty$ to yield a finite, non-zero temperature for all types of stationary motion in all dimensions, $4+1$ and greater.

\subsubsection{Odd dimensions $d\geq 5$} \label{odd dimensions d>5}
Firstly, consider odd dimensions $d\geq 5$. The response function is given by \eqref{RF d dim odd},
\begin{subequations}
    \begin{align} 
    \F_d(E) &= \mathcal{A}_d(E) + \mathcal{B}_d(E), \label{F = A+B}\\
    \mathcal{A}_d(E)&= 2(-1)^{(d-1)/2}k_d \intoinf ds \sin(Es) \Bigg( \frac{1}{\big( - \Delta \mathsf{x}^2(s) \big)^{(d-2)/2}} -\sum_{n=1}^{(d-1)/2} \frac{b_{2n}}{s^{d-2n}} \Bigg), \label{A(E)}\\
     \mathcal{B}_d(E) &= 2\pi k_d \Theta(-E) \sum_{n=1}^{(d-1)/2} \frac{(-1)^{d-n}}{(d-2n-1)!}b_{2n} E^{d-2n-1}, \label{B(E)} 
\end{align}
\end{subequations}
where $k_d=\frac{\Gamma(d/2-1)}{4\pi^{d/2}}$ and $b_{2n}\in\mathbb{R}$ are defined in \eqref{small s expansion odd}.

The expression \eqref{B(E)} is a sum of even powers of $E$ and therefore, as $E\to 0$,
\begin{equation} \label{B small E}
    \mathcal{B}_d(E) = -2(-1)^{(d-1)/2}\pi k_d b_{d-1} \Theta(-E) + O(E^2).
\end{equation}

We can rewrite \eqref{A(E)} as 
\begin{align}
    \mathcal{A}_d(E)&=2(-1)^{(d-1)/2}k_d E \intoinf ds \frac{\sin(Es)}{Es} \Bigg( \frac{s}{\big( - \Delta \mathsf{x}^2(s) \big)^{(d-2)/2}} -\sum_{n=1}^{(d-3)/2} \frac{b_{2n}}{s^{d-2n-1}} - b_{d-1} \Bigg) \notag \\
     &= 2(-1)^{(d-1)/2}k_d E \intoinf ds \frac{\sin(Es)}{Es} \Bigg( \frac{s}{\big( - \Delta \mathsf{x}^2(s) \big)^{(d-2)/2}} -\sum_{n=1}^{(d-3)/2} \frac{b_{2n}}{s^{d-2n-1}} \Bigg)  \notag \\ 
     & \ \ \ \ - (-1)^{(d-1)/2}\pi k_d b_{d-1}\sgn E,  \label{RF odd d dim sec int}
\end{align}
where we have used the standard integral 
\begin{equation}
    \int_0^\infty \frac{\sin(Es)}{Es} = \frac{\pi}{2|E|}.
\end{equation}
By \eqref{small s expansion odd}, the expression in parentheses on the second line of \eqref{RF odd d dim sec int} is $b_{d-1} + O(s^2)$ as $s\to 0$. The sum is $O(1/s^2)$ as $s\to\infty$. Since $\frac{1}{\left|\Delta\mathsf{x}^2(s)\right|} \leq \frac{1}{s^2}$ for all types of stationary motion, the slowest $s/\bigl(-\Delta \mathsf{x}^2(s) \bigr)^{(d-2)/2}$ can decay as $s\to\infty$ is as $O(1/s^{d-3})$. Therefore, for $d\geq 5$, the function $s/\bigl(-\Delta \mathsf{x}^2(s) \bigr)^{(d-2)/2}$ decays at least as fast as $O(1/s^2)$ as $s\to \infty$ and thus the expression in parentheses on the second line of \eqref{RF odd d dim sec int} is integrable. Hence, by dominated convergence, \eqref{RF odd d dim sec int} becomes
\begin{align} \label{RF odd d dim sec int small E}
    \mathcal{A}_d(E) &=2(-1)^{(d-1)/2}k_d E \intoinf ds  \Bigg( \frac{s}{\big( - \Delta \mathsf{x}^2(s) \big)^{(d-2)/2}} -\sum_{n=1}^{(d-3)/2} \frac{b_{2n}}{s^{d-2n-1}} \Bigg)  \notag\\
    & \ \ \ \ - (-1)^{(d-1)/2}\pi k_d b_{d-1}\sgn E + o(E).
\end{align}
Combining \eqref{F = A+B}, \eqref{B small E} and \eqref{RF odd d dim sec int small E}, we obtain
\begin{align}
    \F_d(E) &= 2(-1)^{(d-1)/2}k_d E \intoinf ds  \Bigg( \frac{s}{\big( - \Delta \mathsf{x}^2(s) \big)^{(d-2)/2}} -\sum_{n=1}^{(d-3)/2} \frac{b_{2n}}{s^{d-2n-1}} \Bigg) \notag\\
    & \ \ \ + (-1)^{(d+1)/2} \pi k_d b_{d-1}  + o(E),
\end{align}
where we have used $|E| + 2E\Theta(-E) = E$. The small gap expansion of the response function is of the form \eqref{RF small E} and therefore the temperature is finite and non-zero in the small gap limit.

\subsubsection{Even dimensions $d\geq 6$}
Secondly, consider even dimensions $d\geq 6$. The response function is now given by \eqref{RF d dim even},
\begin{subequations}
    \begin{align}
    \F_d(E) &= \mathcal{C}_d(E) + \mathcal{D}_d(E), \label{F=C+D}\\
    \mathcal{C}_d(E) &= 2k_d\intoinf ds \cos(Es) \Bigg( \frac{1}{\big( \Delta \mathsf{x}^2(s) \big)^{(d-2)/2}} - \sum_{n=1}^{(d-2)/2} \frac{a_{2n}}{s^{d-2n}}  \Bigg),  \label{C(E)}\\
    \mathcal{D}_d(E)&=- 2\pi k_d \Theta(-E) \sum_{n=1}^{(d-2)/2} \frac{(-1)^{d/2+n}}{(d-2n-1)!}a_{2n} E^{d-2n-1}, \label{D(E)}
\end{align}
\end{subequations}
where again $k_d = \frac{\Gamma(d/2-1)}{4\pi^{d/2}}$ and $a_{2n}$ are defined in \eqref{small s expansion even}.

The expression \eqref{D(E)} is a sum of odd powers of $E$ and therefore, as $E\to0$,
\begin{equation} \label{D small E}
    D_d(E) = 2\pi k_d a_{d-2}\Theta(-E)E + O(E^3).
\end{equation}

Consider $C_d(E)$ given in \eqref{C(E)}. By \eqref{small s expansion even}, the expression in parentheses is $O(s^2)$ as $s\to0$. The sum is $O(1/s^2)$ as $s\to\infty$. Following a similar argument as given in Section \ref{odd dimensions d>5}, since $d\geq 6$, then as $s\to\infty$, the function $1/\big( \Delta \mathsf{x}^2(s) \big)^{(d-2)/2}$ decays at least as fast as $O(1/s^4)$. Therefore, the expression in parentheses in \eqref{C(E)} is integrable and hence by dominated convergence,
\begin{subequations}
    \begin{align}
        C_d(E) &= C_d(0) + o(1), \label{C(E) = C(0) + o(1)} \\
        C_d(0) &= 2k_d\intoinf ds\Bigg( \frac{1}{\big( \Delta \mathsf{x}^2(s) \big)^{(d-2)/2}} - \sum_{n=1}^{(d-2)/2} \frac{a_{2n}}{s^{d-2n}}  \Bigg). \label{C(0)}
    \end{align}
\end{subequations}

To find the next-to-leading order contribution to \eqref{C(E)}, consider
\begin{align}
    C_d(E) - C_d(0) &= 2k_dE^2 \int_0^\infty ds \left( \frac{\cos(Es)-1}{E^2s^2} \right)\Bigg( \frac{s^2}{\big( \Delta \mathsf{x}^2(s) \big)^{(d-2)/2}} - \sum_{n=1}^{(d-4)/2} \frac{a_{2n}}{s^{d-2n-2}} - a_{d-2} \Bigg) \notag\\ 
    &= 2k_dE^2 \int_0^\infty ds \left( \frac{\cos(Es)-1}{E^2s^2} \right)\Bigg( \frac{s^2}{\big( \Delta \mathsf{x}^2(s) \big)^{(d-2)/2}} - \sum_{n=1}^{(d-4)/2} \frac{a_{2n}}{s^{d-2n-2}} \Bigg)  + \pi k_d a_{d-2} |E| \label{C(E)-C(0)}
\end{align}
where on the first line we have multiplied and divided by $(Es)^2$ in the integrand, and on the second line we have used the standard integral
\begin{equation}
    \int_0^\infty \frac{ \cos(az)-1}{a^2z^2} = -\frac{\pi}{2|a|}.
\end{equation}
The function in the first pair of parentheses on the second line of \eqref{C(E)-C(0)} has the pointwise limit $-\frac{1}{2}$ and is bounded in absolute value by an $E$-independent constant. From the discussion above \eqref{C(E) = C(0) + o(1)}, it follows that $s^2/\big( \Delta \mathsf{x}^2(s) \big)^{(d-2)/2}$ decays at least as fast as $O(1/s^2)$ as $s\to \infty$. The sum in the second pair of parentheses is also $O(1/s^2)$ as $s\to \infty$. Therefore, the function in the second pair of parentheses is integrable and thus, by dominated convergence, we obtain
\begin{equation} \label{C(E) = C(0) + |E|}
    C_d(E) = C_d(0) + \pi k_d a_{d-2} |E| + O(E^2).
\end{equation}

Combining \eqref{F=C+D}, \eqref{D small E}, \eqref{C(0)} and \eqref{C(E) = C(0) + |E|}, we find
\begin{equation}
    \F_d(E) = 2k_d\intoinf ds\Bigg( \frac{1}{\big( \Delta \mathsf{x}^2(s) \big)^{(d-2)/2}} - \sum_{n=1}^{(d-2)/2} \frac{a_{2n}}{s^{d-2n}}  \Bigg) + \pi k_d a_{d-2} E + O(E^2),
\end{equation}
where we have used $|E| + 2E\Theta(-E) = E$. The small gap expansion is of the form \eqref{RF small E} and therefore the temperature is finite and non-zero in the small gap limit. 

\section{Conclusions}
\label{sec:conclusions}

In this paper, we addressed the response of a UDW detector in the Minkowski vacuum of a massless scalar field in two complementary two-parameter families of stationary motion: 
circular motion, generated by a spatial rotation and a time translation, 
and drifted Rindler motion, generated by a boost and a spatial translation. 
We showed that these two-parameter families of motion can be 
smoothly deformed to each other through a one-parameter family known as parator motion, 
generated by a null rotation and a timelike translation, 
and we observed that the deformation has a sense of uniqueness as the unique real analytic continuation in the parameters of the motion. We then proceeded to show that 
this deformation underlies several observations made in the literature about the detector's response in limiting regimes of circular motion and drifted Rindler motion, including the ultrarelativistic limit of circular motion. 
We also established analytic results regarding the detector's response in drifted Rindler motion in 
$2+1$ and $3+1$ spacetime dimensions in several asymptotic regimes, comparing the results to the corresponding regimes of circular motion. In terms of an effective temperature seen by the detector, defined by the detailed balance relation between excitations and de-excitations, we found that the drifted Rindler temperature remains bounded when the detector's energy gap is large but the proper acceleration is fixed. This is in 
contrast to the circular motion temperature, which can be arbitrarily large in this limit. 

A puzzle that motivated our work is that for circular motion in $2+1$ dimensions the effective temperature is much smaller than the linear acceleration Unruh temperature when the detector's energy gap is small~\cite{Bierman2020}, 
and the potential relevance of this phenomenon for analogue spacetime proposals to observe the circular motion Unruh effect \cite{BEC,Bunney:2023ude}. 
We showed that among all types of stationary motion in spacetime dimensions $2+1$ and higher, 
this phenomenon is unique to $2+1$ dimensions and therein to circular motion and to parator motion. We found that the mathematical reason for this phenomenon is the weak decay of the Wightman function along the detector's trajectory. 
As an intermediate step in the analysis, we presented the detector's response in arbitrary stationary motion in spacetime dimensions $2+1$ and higher as an integral formula without distributional singularities, 
generalising the formulae obtained in dimensions up to $5+1$ from the instantaneous transition rate analysis in~\cite{Loukoregspatialprofile,Hodgkinson_2012}. 

Throughout this paper, we considered the detector in the limit of long interaction and weak coupling within first-order perturbation theory. In this limit, the detector's response is stationary, and the detector's response function is the Fourier transform of the field's Wightman function over the detector's full worldline. The small gap behaviour of the response, determined by the decay of the Wightman function at early and late proper times, is thus sensitive to the assumption that the detector operates at arbitrarily early and late times. 
In experimental settings, where interaction duration is limited by technological and budgetary constraints, the small gap behaviour might hence be different, and this difference could be significant in proposals to observe the circular motion Unruh effect in a $(2+1)$-dimensional analogue spacetime system \cite{Retzker:2007vql,BEC,Lochan:2019osm,Bunney:2023ude}. 
We intend to address this question in a forthcoming paper~\cite{waitingforunruh2}.

\section*{Acknowledgments}

We thank Rick Perche for helpful comments on an early version of the manuscript, Chris Fewster for helpful comments on asymptotics, 
Silke Weinfurtner and other members of the Nottingham Gravity Laboratory for numerous helpful interactions, and an anonymous referee for helpful comments. 
The work of JL was supported by United Kingdom Research and Innovation Science and Technology Facilities Council [grant numbers ST/S002227/1, ST/T006900/1 and ST/Y004523/1]. 
For the purpose of open access, the authors have applied a
CC BY public copyright licence to any Author Accepted
Manuscript version arising.

\appendix

\section{Stationary response function in $d$ dimensions} \label{RF derivation}
In this appendix, we calculate the stationary response function associated with a detector in stationary motion in an arbitrary dimension $d\geq 3$. We work in the long-time interaction limit as defined in Section \ref{Section stationary RF}.

The stationary response function is defined as
\begin{equation} \label{appendix RF}
    \F_d(E) = k_d \intinf ds \frac{e^{-iEs}}{\big( \Delta \mathsf{x}^2(s-i\epsilon) \big)^{(d-2)/2}}
\end{equation}
where $k_d = \frac{\Gamma(d/2-1)}{4\pi^{d/2}}$, and the distributional $\epsilon \rightarrow 0^+ $ limit is understood. In odd dimensions, the denominator has the phases $i^{d-2}$ and $(-i)^{d-2}$ when $s>0$ and $s<0$, respectively. The reciprocal of the spacetime interval in \eqref{appendix RF} has the expansion
\begin{subequations}
    \begin{numcases}{ \frac{1}{(\Delta x^2(s-i\epsilon))^{(d-2)/2}}=} 
        \displaystyle{\sum_{n=1}^{(d-2)/2} \frac{a_{2n}}{(s-i\epsilon)^{d-2n}} + O(1) \text{ as } s-i\epsilon \rightarrow 0,} & \text{for $d$ even} \label{small s expansion even}  \\
        \displaystyle{i^{2-d} \sum_{n=1}^{(d-1)/2} \frac{b_{2n}}{(s-i\epsilon)^{d-2n}} + O(s-i\epsilon) \text{ as } s-i\epsilon \rightarrow 0,} & \text{for $d$ odd} \label{small s expansion odd},
\end{numcases}
\end{subequations}
where $a_{2n}$ and $b_{2n}$ are real-valued coefficients that depend on the spacetime dimension $d$. Note that $a_2=(-1)^{d/2+1}$ and $b_2=1$.

Firstly, let $d$ be even. Adding and subtracting the small $s$ expansion \eqref{small s expansion even} from the Wightman function within the integral in \eqref{appendix RF}, we find
\begin{align} 
    \F_d(E) &= k_d \intinf ds e^{-iEs} \Bigg( \frac{1}{\big( \Delta \mathsf{x}^2(s-i\epsilon) \big)^{(d-2)/2}} - \sum_{n=1}^{(d-2)/2} \frac{a_{2n}}{(s-i\epsilon)^{d-2n}}  \Bigg) 
    \notag 
    \\
    &\hspace{3ex}
    + k_d \sum_{n=1}^{(d-2)/2} a_{2n} \intinf ds \frac{e^{-iEs}}{(s-i\epsilon)^{d-2n}},
\label{RF d dim subtracted}
\end{align}
where we have swapped the order of summation and integration in the last term.

The second integral in \eqref{RF d dim subtracted} can be evaluated using contour integration and the residue theorem to give
\begin{equation}
\intinf ds \frac{e^{-iEs}}{(s-i\epsilon)^{d-2n}} = 2\pi i \Theta (-E) \frac{(-iE)^{d-2n-1}}{(d-2n-1)!}.
\end{equation}
In the first term in \eqref{RF d dim subtracted}, we can take the $\epsilon \rightarrow 0^+$ limit under the integral, justified by the expansion \eqref{small s expansion even} and by the large $s$ fall-off of each term in the integrand. Hence, the response function becomes
\begin{align} \label{RF d dim even}
    \F_d(E) &= 2k_d\intoinf ds \cos(Es) \Bigg( \frac{1}{\big( \Delta \mathsf{x}^2(s) \big)^{(d-2)/2}} - \sum_{n=1}^{(d-2)/2} \frac{a_{2n}}{s^{d-2n}}  \Bigg) \notag \\
    &- 2\pi k_d \Theta(-E) \sum_{n=1}^{(d-2)/2} \frac{(-1)^{d/2+n}}{(d-2n-1)!}a_{2n} E^{d-2n-1},
\end{align}
where we have used the fact that $\big( \Delta \mathsf{x}^2(s) \big)^{(d-2)/2}$ is even in $s$ when $d$ is even. The first term in the sum is the response of an inertial detector and the rest is the non-inertial correction.

Now let $d$ be odd. After adding and subtracting the small $s$ expansion \eqref{small s expansion odd} from the Wightman function within the integral, \eqref{appendix RF} becomes
\begin{align}
    \F_d(E) &= k_d \intinf ds e^{-iEs} \Bigg( \frac{1}{\big( \Delta \mathsf{x}^2(s-i\epsilon) \big)^{(d-2)/2}} - i^{2-d}\sum_{n=1}^{(d-1)/2} \frac{b_{2n}}{(s-i\epsilon)^{d-2n}} \Bigg) \\
    &+ i^{2-d} k_d \sum_{n=1}^{(d-1)/2} b_{2n} \intinf ds \frac{e^{-iEs}}{(s-i\epsilon)^{d-2n}}.
\end{align}
Proceeding as above, the response function is
\begin{align} \label{RF d dim odd}
    \F_d(E) &= 2(-1)^{(d-1)/2}k_d \intoinf ds \sin(Es) \Bigg( \frac{1}{\big( - \Delta \mathsf{x}^2(s) \big)^{(d-2)/2}} -\sum_{n=1}^{(d-1)/2} \frac{b_{2n}}{s^{d-2n}} \Bigg) \notag\\
    &+ 2\pi k_d \Theta(-E) \sum_{n=1}^{(d-1)/2} \frac{(-1)^{d-n}}{(d-2n-1)!}b_{2n} E^{d-2n-1},
\end{align}
where we have used the oddness of $\big( \Delta \mathsf{x}^2(s) \big)^{(d-2)/2}$ to write the integral in terms of $\bigl(- \Delta \mathsf{x}^2(s) \bigr)^{(d-2)/2}>0$ over $s>\infty$.
As in even dimensions, the first term in the sum is the response of an inertial detector and the rest is the non-inertial correction.

\section{Complex singularities of the drifted Rindler Wightman function} \label{sing structure SECTION}

In this appendix we locate and analyse the zeroes of the function 
\begin{equation}
    g(z) = \sinh^2\!z-v^2z^2 , 
\end{equation}
where $z$ is a complex variable and $0<v<1$. 
These zeroes are used in the main text for evaluating the drifted Rindler response function. 

\subsection{Location of the zeroes}

It suffices to consider the zeroes with $\Im z \ge0$. 

Firstly, the only real zero of $g$ is $z=0$, since $\sinh^2z > z^2$ for real nonzero~$z$. 

Secondly, consider the purely imaginary zeroes of $g$. 
Writing $z = i \alpha$, where $\alpha>0$, we find that these zeroes come from the positive solutions to  
\begin{equation} \label{PI zeroes}
    \frac{\sin^2 \! \alpha}{\alpha^2} = v^2 .
\end{equation}
In the interval $(0,\pi)$ there is exactly one solution, which we denote by $\alpha_0$, 
and we parametrise $v$ in terms of $\alpha_0$ as 
$v=\frac{\sin\alpha_0}{\alpha_0}$. 
This is the only solution for $v$ sufficiently close to unity. 
As $v$ decreases, new solutions appear when $v^2$ equals a local maximum value of the function $\frac{\sin^2 \! \alpha}{\alpha^2}$. 
There are finitely many solutions for each~$v$.  
We enumerate the solutions as $\{\alpha_k \}_{k=0}^N$, 
where 
\begin{subequations}
\label{eq:realalphas-enumeration}
\begin{align}
&0 < \alpha_0 < \pi <\alpha_1< \alpha_2 
< \cdots < n \pi < \alpha_{2n-1} < \alpha_{2n} < (n+1) \pi 
\ \ \  \text{for}\ N = 2n, 
\\
&0 < \alpha_0 < \pi <\alpha_1< \alpha_2 
< \cdots < n \pi < \alpha_{2n-1} < (n+1) \pi 
\ \ \ \text{for}\ N = 2n+1, 
\end{align}
\end{subequations}
$n=0,1,2,\ldots$, and all the solutions are simple zeroes of~$g$,  
except that $\alpha_{2n-1}$ is a double zero of $g$ when $N = 2n+1$ with $n\ge1$. 
The double zero occurs when $v^2$ equals a local maximum value of $\frac{\sin^2 \! \alpha}{\alpha^2}$. 

Thirdly, consider the zeroes of $g$ with a positive imaginary part but also a possibly nonvanishing real part. 
We factorise $g$ as 
\begin{subequations}
\begin{align}
    g(z) &= g_+(z)g_-(z), \\
    g_\eta(z) &= \sinh z +\eta vz, 
\end{align}
\end{subequations}
where $\eta = \pm1$. Writing $z=i(\alpha+i\beta)$, where $\alpha>0$ and $\beta \in \mathbb{R}$, 
and decomposing $g_\eta(i\alpha -\beta)$ into its real and imaginary parts, 
we find that the zeroes of $g_\eta$ have imaginary part at the 
positive zeroes of the functions 
\begin{equation}
\label{eq:f-eta-def}
f_\eta(\alpha) =  \cos\alpha 
\, \sqrt{v^2\frac{\alpha^2}{\sin^2\!\alpha}-1} \> 
    + \eta v \arccosh \! \Bigg(v\Bigg| \frac{\alpha}{\sin\alpha} \Bigg|\Bigg) , 
\end{equation}
where 
$f_+$ is defined for $- v \le \frac{\sin\alpha}{\alpha} <0$ 
and 
$f_-$ is defined for $0 < \frac{\sin\alpha}{\alpha} \le v$. 
$f_+$ has come from $g_+$ and $f_-$ has come from $g_-$. 
For each zero of $f_\eta$, the corresponding zero or zeroes of $g_\eta$ are given by $z = i(\alpha+ i\beta)$ 
where 
\begin{equation} \label{betas}
    \beta = \pm \arccosh \!\Bigg(v\Bigg|\frac{\alpha}{\sin\alpha} \Bigg|\Bigg).
\end{equation}

The zeroes of $g$ can now be found by an elementary analysis of \eqref{eq:f-eta-def} and~\eqref{betas}. 
Each strip $n\pi < \Im z< (n+1)\pi$ that contains purely imaginary zeroes, enumerated in \eqref{eq:realalphas-enumeration}, 
contains no other zeroes. 
Each strip $n\pi < \Im z< (n+1)\pi$ that does not contain purely imaginary zeroes contains exactly one pair of zeroes, which are simple and have nonvanishing and opposite real parts. 

\subsection{Complex zeroes at $v\to1$}

We provide here estimates for the complex zeroes as $v\to1$, used in the ultrarelativistic limit analysis in Section~\ref{sec:3+1ultrarel}. 

For $v$ sufficiently close to~$1$, the only purely imaginary zero is $i\alpha_0$. 
We choose a constant $v_0$ for which this holds, 
and we now assume throughout that $v_0 \le v < 1$.   
As noted above, the other complex zeroes then occur in pairs, 
one pair in each strip $n\pi < \Im z< (n+1)\pi$ with $n=1,2,\ldots$. Writing the zeroes 
in the strip $n\pi < \Im z< (n+1)\pi$ as $i(\alpha_n \pm i\beta_n)$, $n=1,2,\ldots$, where $\beta_n>0$, 
an elementary analysis of \eqref{eq:f-eta-def} shows that $n\pi < \alpha_n < (n+\frac12)\pi$ and each $\alpha_n$ is a decreasing function of~$v$. Consideration of the $v\to1$ limit of \eqref{eq:f-eta-def} then shows that there exists a positive numerical constant~$c_1$, independent of~$n$, such that $n\pi + c_1 < \alpha_n < (n+\frac12)\pi$, 
and further that 
$(n+\frac12)\pi - \frac{\ln(n+1)}{2\pi n}\bigl(1 + o(1)\bigr)  < \alpha_n < (n+\frac12)\pi$ as $n\to\infty$, where the $o(1)$ error term is uniform in~$v$. 
From \eqref{betas} it then follows that there exists a positive numerical constant $c_2$ such that $c_2 n < \beta_n\tan\alpha_n$, for all $v_0 \le v<1$ and $n=1,2,\ldots$. 
This will be used to bound $\F_\comp^\corr(E)$ \eqref{3+1 comp RF} in 
Section~\ref{sec:3+1ultrarel}.

\section{3+1 small gap response function for circular and drifted Rindler motion} \label{APP small gap RF}
In this appendix, we calculate the $E\to 0$ expansion of the stationary response function for circular and drifted Rindler motion in $3+1$ dimensions to linear order in $E$.

\subsection{Circular motion}
Consider circular motion in 3+1 dimensions. Using \eqref{CM worldlime} and \eqref{rfd=4}, the non-inertial correction to the response function is
\begin{equation} \label{CM CORR RF}
    \F^\corr(E) = \frac{1}{4\pi^2\gamma v R}\int_0^\infty dz \cos\left( \tfrac{2ER}{\gamma v}z \right)\left( \frac{\gamma^2v^2}{z^2}- \frac{1}{\frac{z^2}{v^2}-\sin^2z} \right),
\end{equation}
where we have changed variables to $z= \frac{\gamma v}{2R} s$. We start by writing \eqref{CM CORR RF} as
\begin{equation} \label{F(a)}
    \F^\corr(a) = \frac{v}{4\pi^2\gamma R} P(a),
\end{equation}
where $a=\tfrac{2R}{\gamma v}E$ and 
\begin{equation} \label{CM RF G(a)}
    P(a) = \int_0^\infty dz \cos(az) \left( \frac{\gamma^2}{z^2} - \frac{1}{z^2\left(1-v^2\frac{\sin^2\!z}{z^2}\right)} \right).
\end{equation}
The expression in parentheses is $\frac{\gamma^4v^2}{3} + O(z^2)$ as $z\to 0$, and it is $\frac{\gamma^2 v^2}{z^2} + O(1/z^4)$ as $z\to \infty$. Therefore, by dominated convergence, $P(a)$ is
\begin{subequations}
\begin{align}
    P(a) &= P(0) + o(1), \\
   P(0) &= \int_0^\infty dz \left(\frac{\gamma^2}{z^2} - \frac{1}{z^2\left(1-v^2\frac{\sin^2\!z}{z^2}\right)} \right)
\end{align}
\end{subequations}
as $a\to 0$.

To find the next-to-leading order term, we first subtract $P(0)$ from both sides of \eqref{CM RF G(a)}. This gives
\begin{align} 
    P(a) - P(0) &= a^2 \int_0^\infty dz \left(\frac{ \cos(az)-1 }{a^2z^2}\right) \left( \gamma^2 - \frac{1}{1-v^2\frac{\sin^2\!z}{z^2}} \right) \notag \\ 
    &=a^2 \int_0^\infty dz\left(\frac{ \cos(az)-1 }{a^2z^2}\right) \left( \gamma^2 - \frac{1}{1-v^2\frac{\sin^2\!z}{z^2}} + (\gamma^2-1) - (\gamma^2-1) \right)\notag \\
    &= (\gamma^2-1) a^2 \int_0^\infty dz \frac{ \cos(az)-1 }{a^2z^2} +a^2\int_0^\infty dz \left(\frac{ \cos(az)-1 }{a^2z^2}\right) \left( 1 - \frac{1}{1-v^2\frac{\sin^2\!z}{z^2}} \right)\notag \\
    &= -\frac{\pi}{2}\gamma^2v^2|a| +a^2\int_0^\infty dz \left(\frac{ \cos(az)-1 }{a^2z^2}\right)\left( 1 - \frac{1}{1-v^2\frac{\sin^2\!z}{z^2}} \right), \label{G(a)-G(0) 4}
\end{align}
where in the second equality we have added and subtracted $\gamma^2-1$ in the second pair of parentheses, in the third equality we have split the integral into two, justified by the integrability of each expression, and in the fourth equality we have used $\gamma^2-1 =\gamma^2v^2$ and the standard integral
\begin{equation} \label{cos integral}
    \int_0^\infty \frac{ \cos(az)-1}{a^2z^2} = -\frac{\pi}{2|a|}.
\end{equation}

The expression in the second pair of parentheses in \eqref{G(a)-G(0) 4} is integrable and the expression in the first pair of parentheses is bounded in absolute value by an $a$-independent constant and has the pointwise limit $-\frac{1}{2}$ as $a\to0$. Therefore, by dominated convergence,
\begin{equation} \label{G(a) small a}
    P(a) = P(0) -\frac{\pi}{2}\gamma^2v^2|a| + O(a^2),
\end{equation}
as $a\to 0$.

Combining \eqref{G(a) small a} with \eqref{F(a)} and using $a=\tfrac{2R}{\gamma v}E$, the small $E$ expansion of $\F^\corr(E)$ is
\begin{equation} 
    \F^\corr(E) = \frac{v}{4\pi^2\gamma R} P(0) - \frac{v^2}{4\pi}|E| + O(E^2).
\end{equation}

Therefore, the small $E$ expansion of the response function $\F(E)$ is
\begin{align}
    \F(E) &= \frac{v}{4\pi^2\gamma R} P(0) - \frac{v^2}{4\pi}|E| -\frac{E}{2\pi}\Theta(-E) + O(E^2) \notag\\
    &= \frac{v}{4\pi^2\gamma R} P(0) -\frac{1}{4\pi}\left(1-\frac{1}{\gamma^2}\sgn E\right)E + O(E^2),
\end{align}
where we have used $v^2 = 1-\frac{1}{\gamma^2}$ and $\sgn(E) + 2\Theta(-E)=1$.

This expansion completes the argument leading to the 3+1 circular motion small gap temperature quoted in equation $(3.9)$ in \cite{Bierman2020}.

\subsection{Drifted Rindler motion}
Consider drifted Rindler motion in 3+1 dimensions. The non-inertial correction to the response function is given by \eqref{RF DR corr},
\begin{equation} \label{RF DR corr APP}
    \F^\corr(E) = \frac{1}{4\pi^2 \gamma  R} \int_0^\infty dz \cos(\tfrac{2ER}{\gamma }z) \Bigg(\frac{\gamma^2}{z^2}-\frac{1}{\sinh^2z-v^2z^2} \Bigg).
\end{equation}
We start by writing \eqref{RF DR corr APP} as 
\begin{equation} \label{Fcorr(b)}
    \F^\corr(b) = \frac{1}{4\pi^2\gamma R}Q(b),
\end{equation}
where $b = \frac{2ER}{\gamma}$ and 
\begin{equation} \label{Q(b)}
    Q(b) = \intoinf dz \cos(bz)\left(\frac{\gamma^2}{z^2}-\frac{1}{z^2\!\left(\frac{\sinh^2z}{z^2}-v^2\right)} \right).
\end{equation}
The expression in parentheses is $-\frac{\gamma^4}{3} + O(z^2)$ as $z\to 0$, and it is $\frac{\gamma^2}{v^2z^2} + O(1/z^4)$ as $z\to \infty$. Therefore, by dominated convergence, $Q(b)$ is
\begin{subequations}
 \begin{align}
    Q(b) &= Q(0) + o(1), \\
    Q(0) &= \intoinf dz \left(\frac{\gamma^2}{z^2}-\frac{1}{z^2\!\left(\frac{\sinh^2z}{z^2}-v^2\right)} \right)
\end{align}   
\end{subequations}
as $b\to 0$.

To find the next-to-leading order term, we subtract $Q(0)$ from both sides of \eqref{Q(b)}. This gives
\begin{align}
    Q(b)-Q(0) &= b^2\intoinf dz \left( \frac{\cos(bz)-1}{b^2z^2}\right)\left(\gamma^2-\frac{1}{\frac{\sinh^2z}{z^2}-v^2} \right) \notag\\
    &= \gamma^2b^2\intoinf dz \frac{\cos(bz)-1}{b^2z^2} -b^2\intoinf dz \left( \frac{\cos(bz)-1}{b^2z^2}\right)\left(\frac{1}{\frac{\sinh^2z}{z^2}-v^2} \right) \notag\\
    &= -\frac{\pi}{2}\gamma^2|b| -b^2\intoinf dz \left( \frac{\cos(bz)-1}{b^2z^2}\right)\left(\frac{1}{\frac{\sinh^2z}{z^2}-v^2} \right), \label{Q(b)-Q(0)}
\end{align}
where in the second equality we have split the integral into two parts, justified by the integrability of each expression, and in the third equality we have used \eqref{cos integral}.

Due to the exponential fall-off of the integrand of the second term in \eqref{Q(b)-Q(0)}, we may use dominated convergence to expand the cosine to all orders under the integral to obtain
\begin{equation} \label{Q(b)-Q(0 2}
    Q(b)-Q(0) = -\frac{\pi}{2}\gamma^2|b| +O(b^2)
\end{equation}
as $b\to 0$.

Combining \eqref{Q(b)-Q(0 2} with \eqref{Fcorr(b)} and using $b=\frac{2ER}{\gamma}$, we find
\begin{equation}
    \F^\corr(E) = \frac{1}{4\pi^2\gamma R}Q(0) - \frac{|E|}{4\pi} + O(E^2)
\end{equation}
as $E\to 0$. Therefore, the small $E$ expansion of the response function is
\begin{align}
    \F(E) &= \frac{1}{4\pi^2\gamma R}Q(0) - \frac{|E|}{4\pi} - \frac{E}{2\pi}\Theta(-E)+ O(E^2) \\
    &=\frac{1}{4\pi^2\gamma R}Q(0) - \frac{E}{4\pi}+ O(E^2)
\end{align}
where we have used $\sgn(E) + 2\Theta(-E)=1$.

\section{$v\rightarrow 0$ and $v\rightarrow 1$ asymptotics of $J(v)$ and $K(v)$} \label{J(v) limits}
In this appendix, we calculate the $v\rightarrow 0$ and $v\rightarrow 1$ asymptotics of the functions $J(v)$ and $K(v)$ defined in \eqref{J(v)} and \eqref{K(v)}, respectively.

\subsection{$J(v)$}
Consider $J(v)$ given by \eqref{J(v)},
\begin{equation} \label{J(v) APP}
    J(v) = \int_0^\infty dz \left( \frac{1}{\gamma z^2} - \frac{1}{\gamma^3(\sinh^2\!z-v^2z^2)} \right),
\end{equation}
where $0\leq v<1$ and $\gamma = (1-v^2)^{-1/2}\geq 1$.

Setting $y=\gamma z$ in \eqref{J(v) APP} 
gives 
\begin{equation} \label{J(v) y}
    J(v) = \int_0^\infty dy \left( \frac{1}{y^2} - \frac{1}{\gamma^4\sinh^2(y/\gamma)-(\gamma^2-1)y^2}  \right).
\end{equation}
For $y>0$, an elementary analysis shows that the integrand in \eqref{J(v) y} is strictly positive and bounded above by the $v$-independent integrable function $\frac{1}{y^2} - \frac{1}{\sinh^2 y}$. 
The $v\to 0$ and $v\to1$ limits can hence be taken under the integral by dominated convergence. 
At $v\to0$, we find 
\begin{equation} 
    J(v) = 1 + o(1), 
\end{equation}
where we have used 
\begin{equation} 
\label{J(v) v=0-integral}
    J(0) = \int_0^\infty dz \left( \frac{1}{z^2} - \frac{1}{\sinh^2\!z} \right)  = 1 . 
\end{equation}
The integral in \eqref{J(v) v=0-integral} can be evaluated by extending the lower limit of the integral to $-\infty$ by evenness, deforming the contour to $z = i\frac{\pi}2 + r$ with $r \in \mathbb{R}$, and using 3.511.8 in \cite{Gradshteyn:1943cpj}. 
At $v\to1$, we find
\begin{align} 
    J(v) &= \int_0^\infty dy \left( \frac{1}{y^2} - \frac{1}{y^2+\frac{y^4}{3}} \right) + o(1)
\notag
\\
&= \frac{\pi}{2\sqrt{3}} + o(1) , 
\label{J(v) elementary integral}
\end{align}
where the last equality follows by evaluating the elementary integral. 

We also note that $J(v)$ is a monotonically decreasing function of~$v$. This follows because the integrand in \eqref{J(v) y} is a monotonically decreasing function of $v$ for $0\leq v <1$ at fixed $y>0$.

\subsection{$K(v)$}

Consider $K(v)$ given by \eqref{K(v)}, 
\begin{equation} \label{K(v) 2}
    K(v) = \frac{4}{\pi^2} \int_0^\infty dz \frac{z}{\sqrt{\sinh^2 \! z-v^2z^2}}, 
\end{equation}
where $0\le v < 1$. 

For the $v\to0$ limit, we observe that when $v\le \frac12$, the integrand in \eqref{K(v) 2} is bounded above by the integrable $v$-independent function 
$z / \! \left(\sinh^2 \!z- \frac14 z^2 \right)^{-1/2}$. 
The $v\to0$ limit can hence be taken by dominated convergence, with the outcome 
\begin{align}
    K(v) = 1 + o(1),
\end{align}
where we have used 3.521.1 in \cite{Gradshteyn:1943cpj} to evaluate the integral in 
\begin{align}
    K(0) = \frac{4}{\pi^2}\int_0^\infty dz \frac{z}{\sinh z} = 1 . 
\end{align}

For the $v\to 1$ limit, we write $K(v)$ \eqref{K(v) 2} as 
\begin{equation} \label{K(v) v->1}
    K(v) = \frac{4}{\pi^2}\int_0^\infty dz \frac{1}{\sqrt{h(z)+ \frac{1}{\gamma^2}}},
\end{equation}
where $h(z) = \frac{\sinh^2 \! z}{z^2} -1$ and $\gamma = (1-v^2)^{-1/2}$. 
Note that $h(z) = \frac{1}{3}z^2 + O(z^4)$ as $z\to 0$. Introducing a positive constant $M$, we can then split the domain of integration in \eqref{K(v) v->1} as
\begin{subequations}
\begin{align}
    K(v) &= K_<(v) + K_>(v), \\
    K_<(v) &= \frac{4}{\pi^2}\int_0^M dz \frac{1}{\sqrt{h(z)+ \frac{1}{\gamma^2}}}, 
    \label{eq:K<def}\\
    K_>(v) &= \frac{4}{\pi^2}\int_M^\infty dz \frac{1}{\sqrt{h(z)+ \frac{1}{\gamma^2}}}.
\end{align}
\end{subequations}

Firstly, consider $K_>(v)$. Expanding the integrand in powers of $\frac{1}{\gamma^2}$, we find
\begin{equation} \label{K>}
    K_>(v) = \frac{4}{\pi^2}\int_M^\infty dz \frac{1}{\sqrt{h(z)}} + O\!\left(\frac{1}{\gamma^2}\right),
\end{equation}
where interchanging the expansion and the integral is justified by dominated convergence.

Secondly, consider $K_<(v)$. In the integrand in \eqref{eq:K<def}, we add and subtract a term in which $h(z)$ is replaced by its leading small $z$ term $\frac13 z^2$, and we regroup the integral as 
\begin{equation} \label{K<}
    K_<(v) = \frac{4}{\pi^2} \int_0^M dz \left(\frac{1}{\sqrt{h(z)+\frac{1}{\gamma^2}}} - \frac{1}{\sqrt{\frac{1}{3}z^2 + \frac{1}{\gamma^2}}} \right) + \frac{4}{\pi^2} \int_0^M dz\frac{1}{\sqrt{\frac{1}{3}z^2 + \frac{1}{\gamma^2}}}.
\end{equation}
In the first integral in \eqref{K<}, the integrand can be written as 
\begin{align}
    \frac{1}{\sqrt{h(z)+\frac{1}{\gamma^2}}} - \frac{1}{\sqrt{\frac{1}{3}z^2 + \frac{1}{\gamma^2}}} 
    &=  \frac{\frac{1}{3}z^2-h(z)}{\sqrt{h(z)+\frac{1}{\gamma^2}}\sqrt{\frac{1}{3}z^2 + \frac{1}{\gamma^2}}\left[ \sqrt{\frac{1}{3}z^2 + \frac{1}{\gamma^2}} +\sqrt{h(z)+\frac{1}{\gamma^2}}\right]}, 
\end{align}
which shows that the integrand is bounded in absolute value by
\begin{equation}
    \frac{h(z) - \frac{1}{3}z^2}{\sqrt{h(z)}\sqrt{\frac{1}{3}z^2} \left[ \sqrt{\frac{1}{3}z^2}+\sqrt{h(z)} \right]}, 
\end{equation}
which is independent of $\gamma$ and integrable over $[0,M]$. By dominated convergence, we hence have 
\begin{equation} \label{K<1}
    \frac{4}{\pi^2} \int_0^M dz \left(\frac{1}{\sqrt{h(z)+\frac{1}{\gamma^2}}} - \frac{1}{\sqrt{\frac{z^2}{3} + \frac{1}{\gamma^2}}} \right) = \frac{4}{\pi^2} \int_0^M dz \left(\frac{1}{\sqrt{h(z)}}- \frac{\sqrt{3}}{z}\right) + o(1) 
\end{equation}
as $\gamma \to\infty$. 

The second integral in \eqref{K<} is elementary and its large $\gamma$ expansion is
\begin{equation} \label{K<2}
    \frac{4}{\pi^2} \int_0^M dz\frac{1}{\sqrt{\frac{1}{3}z^2 + \frac{1}{\gamma^2}}} = \frac{4\sqrt{3}}{\pi^2} \log(\frac{2\sqrt{3}}{3}\gamma) + \frac{4\sqrt{3}}{\pi^2}\log M + O\!\left(\frac{1}{\gamma^2}\right).
\end{equation}
Combining \eqref{K>}, \eqref{K<1} and $\eqref{K<2}$, we find 
\begin{equation} \label{K(v) v->1 2}
    K(v) = \frac{4\sqrt{3}}{\pi^2} \log(\frac{2\sqrt{3}}{3}\gamma)+ \frac{4\sqrt{3}}{\pi^2}\log(M) + \frac{4}{\pi^2}\int_M^\infty dz \frac{1}{\sqrt{h(z)}}+\frac{4}{\pi^2} \int_0^M dz \left(\frac{1}{\sqrt{h(z)}}- \frac{\sqrt{3}}{z}\right) + o(1) 
\end{equation}
as $\gamma \to \infty$. 

The sum of the three individually $M$-dependent terms in \eqref{K(v) v->1 2} is independent of~$M$, 
as can be seen by differentiating the sum with respect to~$M$. To write the sum in an explicitly $M$-independent way, 
we write 
\begin{equation}
    \log M = \int_0^M \frac{dz}{1+z} - \log\!\left(1+\frac{1}{M}\right),
\end{equation}
and we group the sum as
\begin{align}
   &\frac{4}{\pi^2} \int_0^M dz \left(\frac{1}{\sqrt{h(z)}}- \frac{\sqrt{3}}{z(1+z)}\right) + \frac{4}{\pi^2}\int_M^\infty dz \frac{1}{\sqrt{h(z)}} \ - \ \frac{4\sqrt{3}}{\pi^2} \log\left(1+\frac{1}{M}\right) \notag \\
   & \ \ = \frac{4}{\pi^2} \int_0^\infty dz \left(\frac{1}{\sqrt{h(z)}}- \frac{\sqrt{3}}{z(1+z)}\right), 
\end{align}
where in the last equality we have used the $M$-independence of the sum to take the limit $M\to\infty$ termwise.  
Hence, as $v\to 1$ ($\gamma\to\infty)$, 
\begin{equation} \label{K(v) ref}
    K(v) = \frac{4\sqrt{3}}{\pi^2} \log(\frac{2\sqrt{3}}{3}\gamma)+\frac{4}{\pi^2} \int_0^\infty dz \left(\frac{z}{\sqrt{\sinh^2\!z -z^2}}- \frac{\sqrt{3}}{z(1+z)}\right)  + o(1).
\end{equation}

\section{2+1 drifted Rindler ultrarelativistic limit} \label{Append 2+1 ultrarel lim}
In this appendix, we verify the $v\to 1$ ($\gamma \to \infty)$ limit \eqref{RF odd v->1} with fixed $E/a$ for drifted Rindler motion in 2+1 dimensions.

Starting with the expression for the odd part of the response function in 2+1 dimensions as given in \eqref{RFodd}, we make the change of variables $y=\gamma z$ to obtain
\begin{equation} \label{APP F ODD}
     \F^\text{odd}(E) = -\frac{1}{2\pi} \intoinf dy \, \frac{\sin(2\tfrac{E}{a}y)}{\sqrt{\gamma^4\sinh^2(\tfrac{y}{\gamma})-(\gamma^2-1)y^2}},
\end{equation}
where we have used \eqref{DR acceleration torsion}.  

An elementary analysis shows that the integrand in \eqref{APP F ODD} is bounded in absolute value by 
$\bigl|\sin(2\tfrac{E}{a}y)\bigr| y^{-1} {\bigl(1+\frac{1}{3}y^2\bigr)}^{-1/2}$, 
which is independent of $\gamma$ and integrable in~$y$. 
We can hence take the $\gamma\to\infty$ limit in \eqref{APP F ODD} under the integral, with the result 
\begin{equation}
    \F^\odd(E) = -\frac{1}{2\pi} \intoinf dy \, \frac{\sin\big( 2\tfrac{E}{a}y \big)}{y\sqrt{1+\frac{1}{3}y^2}} + o(1). 
\end{equation}
Finally, by the change of variables $x=\frac{y}{\sqrt{3}}$, we have   
\begin{equation}
    \F^\odd(E) = -\frac{1}{2\pi} \intoinf dx \, \frac{\sin\big( 2\sqrt{3}\tfrac{E}{a}x \big)}{x\sqrt{1+x^2}} + o(1) = -\frac{1}{2\pi}G(2\sqrt{3}E/a) + o(1)
\end{equation}
as $\gamma \to \infty$, where the function $G$ is as defined in~\eqref{G(q)}.

\bibliography{references}

\end{document}